\documentclass[10pt]{article}%
\usepackage{mystyle_article}
\usepackage[font=small]{caption}
\usepackage{upgreek}

\newcommand{\heart}{\heartsuit}

\mathchardef\mhyphen="2D

\usepackage{booktabs}
\usepackage{footnote}
\makesavenoteenv{minipage}
\title{\vspace{-1.3cm}Strategic Behavior under Context Misalignment%
\thanks{Pierfrancesco thankfully acknowledges financial support from the Austrian Science Fund (FWF) (P31248-G27) and from MIUR under the PRIN 2017 program (grant number 2017K8ANN4).}
\\
}
\author{%
Pierfrancesco Guarino\thanks{%
	University of Verona (Department of Economics). %
	\textit{E-mail:} \texttt{pf.guarino@hotmail.com}.}
{\ \& }%
Gabriel Ziegler\thanks{%
 	University of Edinburgh (School of Economics). %
 	\textit{E-mail:} \texttt{ziegler@ed.ac.uk}.}%
}
\date{\mydate\today}
\begin{document}
\renewcommand\thmcontinues[1]{Continued}

\maketitle


\begin{abstract}
\noindent We study the behavioral implications of Rationality and Common Strong Belief in Rationality (RCSBR) with contextual assumptions allowing players to entertain misaligned beliefs, i.e., players can hold beliefs concerning their opponents’ beliefs where there is no opponent holding those very beliefs. Taking the analysts’ perspective, we distinguish the infinite hierarchies of beliefs actually held by players (``real types'') from those that are a byproduct of players’ hierarchies (``imaginary types'') by introducing the notion of separating type structure. We characterize the behavioral implications of RCSBR for the real types across all separating type structures via a family of subsets of Full Strong Best-Reply Sets of \cite{Battigalli_Friedenberg_2012}. By allowing misalignment, in dynamic games we can obtain behavioral predictions inconsistent with RCSBR (in the standard framework), contrary to the case of belief-based analyses for static games---a difference due to the dichotomy ``non-monotonic vs. monotonic'' reasoning.

\bigskip



\end{abstract}

\section{Introduction}
\label{sec:introduction}

\subsection{Motivation \& Results}
\label{subsec:motivation_results}

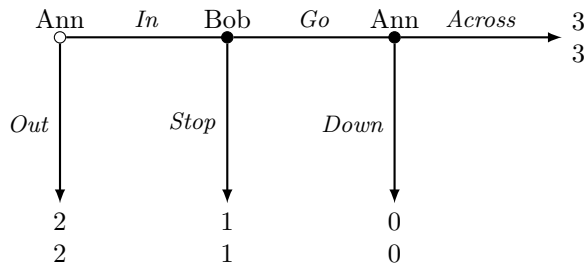
\begin{figure}[H]
	\centering
	\begin{tikzpicture}
		[info/.style={circle, draw, inner sep=1.5, fill=black},
		scale=1.1] 
		\node[info, fill=white] (n_1) at (0,4.5) {};
		\draw[thick, -latex] (n_1) -- node[left] {\small\emph{Out}} (0,2.5) node[below, align=center] {$2$\\$2$};
		\draw[thick, -] (n_1) -- node[above] {\small\emph{In}} (2,4.5);
		\draw (n_1) node[above] {Ann};
		\node[info] (n_2) at (2,4.5) {};
		\draw[thick, -latex] (n_2) -- node[left] {\small\emph{Stop}} (2,2.5) node[below, align=center] {$1$\\$1$};
		\draw[thick, -] (n_2) -- node[above] {\small\emph{Go}} (4,4.5);
		\draw (n_2) node[above] {Bob};
		\node[info] (n_3) at (4,4.5) {};
		\draw[thick, -latex] (n_3) -- node[left] {\small\emph{Down}} (4,2.5)  node[below, align=center] {$0$\\$0$};
		\draw[thick, -latex] (n_3) -- node[above] {\small\emph{Across}} (6,4.5)  node[right, align=center] {$3$\\$3$};
		\draw (n_3) node[above] {Ann};
	\end{tikzpicture}
	\caption{Common-interest centipede.}
	\label{fig:CIC}
\end{figure}

Consider an analyst investigating rational behavior in the common-interest and centipede-like game represented in \autoref{fig:CIC} as initially studied by \citet[Figure 6, p.76]{Battigalli_Friedenberg_2012}. Suppose the observed outcome of the game is $\{In\} \times \{Stop\}$ (i.e., Ann lets Bob decide whether she can move again, but Bob stops the game), where both players receive a payoffs of $1$. The analyst wonders if this outcome is consistent with three baseline assumptions:\footnote{Here, we are purposefully informal to avoid too much notation: formal definitions are---of course---going to be introduced in due course.} ($i$) both players are (sequentially) rational; ($ii$) they reason about their rationality; ($iii$) they try to rationalize the opponent's behavior whenever possible. 

\cite{Battigalli_Siniscalchi_2002} formalize these assumptions via their notion of \emph{Rationality and Common Strong Belief in Rationality}. Building on one of their results, the analyst could take the predictions of \emph{Strong Rationalizability}\footnote{\label{foot:SR}A common name for this solution concept---only hinted in \cite{Pearce_1984}, but adopted in \cite{Battigalli_1997} and subsequent work---is ``Extensive-Form Rationalizability''. Here we employ a terminology introduced in \cite{Battigalli_1999}, which distinguishes this solution concept from other forms of Rationalizability that can be formalized for the analysis of dynamic games in their extensive form (i.e., Initial (or Weak) Rationalizability \emph{\`a la} \cite{Ben-Porath_1997} and Backward Rationalizability \emph{\`a la} \cite{Penta_2015}). For the latter, see also \citet{Perea_2014}, \citet{Battigalli_De_Vito_2021}, and \citet{Catonini_Penta_2022}.} of \citet[Definition 9, p.1042]{Pearce_1984} and \citet[Definition 2, p.46]{Battigalli_1997} as the behavioral implications corresponding to these assumptions. In \autoref{fig:CIC}, this procedure would yield the standard backward (and forward) induction prediction of $\{In\mhyphen Across\} \times \{Go\}$, which rules out the observed outcome. Upon further scrutiny, the analyst realizes that \emph{Strong Rationalizability} only captures the aforementioned baseline assumptions when there is the additional assumption that there are no exogenous restrictions on the reasoning process of the players and this assumption itself is transparent\footnote{That is, common knowledge in the informal sense of the term.} to the players. \cite{Battigalli_Friedenberg_2012} study the implications on behavior robust to any such exogenous, but transparent, restriction on reasoning and characterize it via the family of \emph{Full Strong Best-Reply Sets}.\footnote{\label{foot:fSBRs}The terminology varies. These objects are called ``Extensive-Form Best Response Sets'' in \citet[Definition 9, p.63]{Battigalli_Friedenberg_2012}, whereas they are called ``Full Extensive-Form Best-Reply Sets'' in \citet[Definition 12.29, p.680]{Dekel_Siniscalchi_2015}. We opt for the name ``Full Strong Best-Reply Sets'' by partially following  \citet[Definition 12.29, p.680]{Dekel_Siniscalchi_2015}, but---at the same time---stressing the relation of this solution concept to Strong Rationalizability (and the related terminology, as described in \Sref{foot:SR}).} In \autoref{fig:CIC}, besides the outcome corresponding to Strong Rationalizability, any other Full Strong Best-Reply Set would predict Ann to play $Out$. Thus, even these weaker assumptions rule out the observed outcome and would predict each player to obtain a payoff of at least $2$.

However, the baseline assumptions do not specify whether or not any restriction on reasoning should be transparent between the players. Indeed, relaxing this assumption and characterizing the robust behavioral implications is the main goal of this paper and, after doing so, we show that the observation of  $\{In\} \times \{Stop\}$ \emph{is consistent with the baseline assumptions}, but it requires players to reason in a misaligned manner, which me make precise in a conceptual contribution of this paper.

Going beyond the motivating example and getting slightly more formal, imagine an analyst involved in the study of any strategic interaction with complete information. Being confident concerning the rules of the game and the payoff functions, the analyst contemplates the following scenario concerning the players in the strategic interaction under scrutiny:
\begin{enumerate}[leftmargin=*, label=\arabic*)]
\item they hold beliefs in a specific subset in the set of all possible beliefs they can have, i.e., we are in presence of \emph{contextual assumptions}, which are assumptions on what players consider or do not consider possible;

\item there is the possibility that they can have \emph{misaligned beliefs}, i.e., a player can hold beliefs concerning her opponents' beliefs where actually there is no opponent that holds these beliefs or---in other words---she can be `wrong' concerning the beliefs held by her opponents.
\end{enumerate}
Given this setting,  the analyst would like to obtain the behavioral predictions stemming from a version of the assumption that players are rational and there is common belief in rationality.

Epistemic game theory\footnote{Concerning epistemic game theory, see the survey by \cite{Dekel_Siniscalchi_2015}, or \cite{Perea_2012} and \cite{Battigalli_et_al_forthcoming},  that are two textbooks completely devoted to the topic.} has extensively studied this problem in \emph{absence} of misaligned beliefs for different classes of games and different notions of rationality, as in  \cite{Brandenburger_Dekel_1987}, \cite{Brandenburger_Friedenberg_2010}, and \cite{Battigalli_Friedenberg_2012}.\footnote{See also \cite{Brandenburger_et_al_2021}, who study the implications of finite-order reasoning in dynamic games.}  To obtain these results, the authors of these works performed the following steps:  (1) they fixed a class of games as the focus of their analysis; (2) they established the opportune kind of epistemic type structures as the \emph{theoretical framework} where to formalize the assumptions of interest for their analysis; (3) they defined an appropriate notion of rationality as the \emph{assumption} under scrutiny with the language provided by the aforementioned theoretical framework; (4) they introduced appropriate \emph{modal operators} to capture the forms of interactive reasoning of interest; (5) they derived the \emph{behavioral predictions} corresponding to the---iterated---application of the previously defined modal operator of interest on the appropriate notion of rationality within a given epistemic type structure. In this line of research, the ``contextual assumptions'' element enters into their analysis at the very last stage: in fixing a given epistemic type structure, these authors address the question of what are the behavioral implications that can be obtained when players hold a specific subset within the set of all possible beliefs they can have.\footnote{In their working paper version, \cite{Battigalli_Friedenberg_2012} formalize the idea that `small' type structures can be seen as equivalent to assuming that only certain subsets of beliefs are considered by the players and this itself is transparent between the players.}  Typically, these behavioral predictions take the form of product subsets of the strategy space with specific properties.

As already pointed out, rather crucially, these analyses (and the corresponding characterizations) rely on players being `correct' with respect to the beliefs held by their opponents, i.e., for every infinite hierarchy of coherent beliefs of a player present in an epistemic type structure, there are related infinite hierarchies of beliefs of her opponents in the same epistemic type structure. In other words, the basic assumption of the endeavour is that the infinite hierarchies of coherent beliefs under scrutiny are transparent between the players. Thus, it seems most natural to investigate what are the behavioral predictions that a modeler---as an outside observer---can obtain when players can have misaligned beliefs. As a matter of fact, the importance of the present venture falls into the so-called \emph{Wilson's doctrine} as put forth in \cite{Wilson_1987}: here, too, we are relaxing the common knowledge assumptions present in a game-theoretical analysis.

We perform our analysis by focusing on dynamic games, extending in particular  \cite{Battigalli_Friedenberg_2012}. With respect to the five stylized points above, these authors (1) focus on dynamic games, (2) employ epistemic type structures with belief functions  opportunely defined to capture conditional beliefs, (3) study the notion of sequential rationality, (4) work with the \emph{strong belief} operator of \citet[Section 3]{Battigalli_Siniscalchi_2002}, (5) and---finally---characterize the notion of Rationality and Common Strong Belief in Rationality across all epistemic type structures via the notion of Full Strong Best-Reply Sets. As a matter of fact, the choice of studying dynamic games somewhat bypassing static games is a conscious decision on our part. Indeed, by focusing on dynamic games (in their extensive form representation), we have static games as a degenerate case, which in turn allows us to uncover that the crucial distinction between these classes of games with respect to our analysis does not lie in the different representation, but rather in the dichotomy ``non-monotonic vs. monotonic'' reasoning which arises at the level of modal operators used to perform the epistemic analysis---a point which we address in detail in \Sref{sec:non-monotonicity}. 

Our primitive notion is that of infinite hierarchies of coherent beliefs, as formally introduced in \Sref{subsec:hierarchies}. Hence, we work with the canonical hierarchical structure, i.e., the type structure that is comprised of \emph{all} those infinite hierarchies of coherent beliefs. Starting from this structure, we define a notion of state space (in \Mref{def:state_space})  as a subset of the \emph{universal state space} induced by the canonical hierarchical structure. This is a crucial point, because this definition allows for a state space to be non-belief-closed (as in \Mref{def:non-belief-closed}), which is exactly how we capture misaligned beliefs, whereas typically the literature has focused on belief-closed state spaces, thus dropping the possibility of studying misaligned beliefs.

In  \Sref{sec:derived_objects}, we illustrate how it is possible to obtain a---somewhat na{\"i}ve---form of `closure' of a non-belief-closed state-space, to then perform the standard analysis: the idea would simply be to take a belief-closed state space that contains all the infinite hierarchies in the original non-belief-closed-state-space. However, such a `closure' would introduce spurious information, since those previously misaligned beliefs would now find support in the infinite hierarchies of coherent beliefs just introduced to `close' the state space, and would then `force' players to reason about types not actually contemplated by them, but only by the analyst. To explicitly avoid these side-effects, we introduce the novel notion of \emph{separating closure} (as in \Mref{def:closure}) to keep track of and distinguish the types in a state space that are \emph{actual} infinite hierarchies of coherent beliefs held by the players---called ``real types''---from those that are a byproduct of other infinite hierarchies of coherent beliefs---called ``imaginary types''. This allows us to formalize, for every player involved in a strategic interaction, the derived notion of \emph{separating type structure} (as in \Mref{def:separating_type_structure}), which is the natural completion in the context of type structures of the notion of separating closure. Building on these objects, we formalize the notion of \emph{player dependent real Rationality and Common Strong Belief in Rationality} (as in \Mref{def:player_RCSBR}), which  essentially captures all those real types in a separating type structure that satisfy Rationality and Common Strong Belief in Rationality, along with the corresponding notion from the perspective of an analyst (as in \Mref{def:analyst_RCSBR}).

Our main results, namely, \Mref{prop:characterization_static_1} and \Mref{prop:characterization_static_2}, fully answer the question of what are the behavioral implications of Rationality and Common Strong Belief in Rationality allowing for the possibility of misaligned beliefs by employing arbitrary separating type structures. In order to see these results in perspective, three points need to be emphasized. First of all, in comparison to \cite{Battigalli_Friedenberg_2012}, we operate a conceptual shift in the treatment of Full Strong Best-Reply Sets by also focusing on a related notion entrenched in the perspective of a single player. In particular, to achieve this result, we introduce the notion of \emph{player-specific Full Strong Best-Reply Set} (as in \Mref{def:player-specific_FSBRS}), which corresponds to a subset of a given player's strategy set such that there exists a product of subsets of the other players' strategies with the property that the product of all those subsets is a Full Strong Best-Reply Set in its own rights. As a result, the product of a collection of player-specific Full Strong Best-Reply Sets, one for every player, is not necessarily a Full Strong Best-Reply Set. Secondly, we introduce the novel notion of Misaligned Full Strong Best-Reply Set (as in \Mref{def:MFSBRS}), that---as the name suggests---stems from that of Full Strong Best-Reply Set. The key issue here lies in the fact that those strategies of a given player that belong to a Misaligned Full Strong Best-Reply Set are subsets of her part of a Full Strong Best-Reply Set and inherit a \emph{fullness} criterion from this very Full Strong Best-Reply Set. As a matter of fact, for our purposes, we also extend this notion to a corresponding player-specific version (as in \Mref{def:player-specific_MFSBRS}). In third place, the novel epistemic apparatus described in the previous paragraph allows to naturally obtain a fine-grained taxonomy of separating type structures according to the properties of these objects with respect to  the real and imaginary types of the various players. In particular, given a player, we consider her corresponding separating type structure \emph{degenerate} if it does not contain imaginary types  of that very player (otherwise being non-degenerate); also, we consider a family of separating type structure---one for every player---\emph{common} if the real and imaginary type spaces are the same according to \emph{every} separating type structure in the family, for every player (otherwise being non-common). As a result, overall we obtain the following taxonomy: separating type structures can be non-degenerate and non-common, non-degenerate and common, degenerate and non-common, and degenerate and common.

The importance of these three points lies in the fact that our characterization results in \Mref{prop:characterization_static_1} and \Mref{prop:characterization_static_2} link (player-specifc) Misaligned Full Strong Best-Reply Sets---along with  (player-specific) Full Strong Best-Reply Sets---to the various possibilities covered in our taxonomy of separating type structures. In particular, in \Mref{prop:characterization_static_1}, we show that, if there are real types that satisfy Rationality and Common Strong Belief in Rationality, then the resulting behavioral predictions associated to those types coincide with one of those forms of subsets, where the actual form depends on the nature of the separating type structure (as captured by our taxonomy). On the contrary, in  \Mref{prop:characterization_static_2}, we show that  starting with such a form of subset of strategies, it is always possible to construct a separating type structure of a certain nature (according to our taxonomy), where the behavioral predictions corresponding to her real types satisfying Rationality and Common Strong Belief in Rationality coincide with this particular subset. Thus, these two propositions together constitute the usual necessary and sufficient epistemic conditions for behavioral implications.\footnote{\citet[Section 12.3.2]{Dekel_Siniscalchi_2015} discuss how necessary epistemic conditions can be interpreted. Already \citet[Section 7.h]{Aumann_Brandenburger_1995} elaborate on related issues for necessary epistemic conditions of Nash equilibrium.} Hence, in order to appreciate the cutting power of the framework introduced, for example, it turns out that the epistemic characterization of Full Strong Best-Reply Sets obtained in \citet[Theorem 1, pp.66--67]{Battigalli_Friedenberg_2012} is covered as a particular case of \Mref{prop:characterization_static_1} and \Mref{prop:characterization_static_2}, namely, Part (4), where the focus is on the  case that corresponds to separating type structures being both degenerate and common.

\subsection{Related Literature}
\label{subsec:related_literature}

The closest paper related to the present work is \cite{Battigalli_Friedenberg_2012}  with its focus on the relation between the epistemic notion of \emph{Rationality and Common Strong Belief in Rationality} and the existence of a solution concept that characterizes such notion across all possible epistemic type structures. However, in contrast to their analysis, we explicitly take the perspective of an analyst that contemplates players with misaligned beliefs. Thus, with respect to this point, the framework developed here is reminiscent of the notion of \emph{player-specific type structure} sketched in \citet[Section 8.1]{Brandenburger_Friedenberg_2010}. We discuss this point in more detail in \Sref{subsec:player_specfic_structures}. Of course, as can be inferred from \Sref{subsec:motivation_results}, other related papers are \cite{Brandenburger_Dekel_1987} and \cite{Brandenburger_Friedenberg_2010}.

Another interpretation of our analysis is that the analyst considers players entertaining potentially wrong models of the situation they are actually in. With this interpretation in mind, the present paper is related to the literature on learning under misspecification, as in \cite{Esponda_Pouzo_2016}, \cite{Frick_et_al_2020}, \cite{Bohren_Hauser_2021}, or \cite{Fudenberg_et_al_2021}.\footnote{As this is a fast growing research area, we refrain from giving a literature review here. The interested reader is referred to the references in the cited papers.} All these studies differ from our paper along two important dimensions: ($i$) we focus on allowing players being wrong in the way they reason about the strategic environment they are in, whereas these studies focus on players being wrong about how they learn about the strategic situation they are in; ($ii$) fundamentally, whereas in our approach players might be wrong about the endogenous uncertainty they are facing, these studies allow for a misspecification of how players interpret information about exogenous uncertainty. Actually, \cite{Friedenberg_Meier_2017} is closer to our approach along the first dimensions, since in this paper the authors study a misspecification of the type space by the analyst in a Bayesian game. It has to be observed that the authors maintain the assumption that the model of reasoning is transparent between the players (i.e., they only consider belief-closed state spaces of the canonical hierarchical structure built on an exogenous parameter space) and that,  in their model, endogenous uncertainty is resolved by assuming equilibrium play. Finally, \cite{Piermont_Zuazo-Garin_2021} study dynamic games with payoff uncertainty and focus on RCSBR, but allow for (higher-order) disagreement between the players about the state space representing the exogenous uncertainty. In their paper the reasoning process is assumed to be transparent between the players as it is by all the other studies cited in this section.

\subsection{Synopsis}

This paper is structured as follows: in \Sref{sec:background_knowledge} we introduce those (game-theoretical and epistemological) objects that are needed to develop the necessary background knowledge. In \Sref{sec:derived_objects}, we introduce the main tools of our work, namely, separating type structures and real types, that we employ in \Sref{sec:results}, where we state our characterization result. In \Sref{sec:non-monotonicity}, we address the differences arising between belief-based analyses of static games and dynamic games stemming from the dichotomy ``non-monotonic/monotonic'' reasoning.  Finally, in \Sref{sec:discussion}, we discuss various aspects related to this work. All the proofs of the results established in the paper are relegated to \Sref{app:proofs}.

\section{Background Knowledge}
\label{sec:background_knowledge}

\subsection{Game-Theoretical Framework}
\label{subsec:game_theory}

The primitive object of our analysis is a finite dynamic game with perfect recall in its extensive form representation (henceforth, dynamic game), which is a tuple\footnote{For a similar definition, see \citet[Definition 200.1, Chapter 11.1.2]{Osborne_Rubinstein_1994} or \citet[Section 2]{Battigalli_Friedenberg_2012}.}
\begin{equation}
\label{eq:dynamic_ordinal_game}
\Gamma := \la I , (A_j)_{j \in I} , X, (H_j, \mbf{S}_j, S_j, u_j)_{j \in I} \ra,
\end{equation} 
where this definition possibly allows for simultaneous moves.  In \Mref{eq:dynamic_ordinal_game}, $I$ denotes the set of players and $A_i$, with $i \in I$, is the set of \emph{actions} of  player $i$. The set $X$ is the set of \emph{histories}, where a history $x$ is either the empty sequence $\la \varnothing \ra$, alternatively called the \emph{initial history}, or it is a sequence $\la a^1, \dots, a^K \ra$,  where $a^k := (a^{k}_{j})_{j \in I}$ with $a^{k}_i \in A_i$ for every $i \in I$ and for every $1 \leq k \leq K$. We let $A (x) : = \prod_{j \in I} A_j (x)$ denote the set of actions available to the players at history $x$: if $\abs{A_i (x)} \geq 2$, then player $i$ is \emph{active} at history $x$, otherwise she is inactive, where we let $\abs{Y}$ denote the cardinality of an arbitrary set $Y$. The set of histories $X$ is an \emph{arborescence} when endowed with a binary relation capturing the notion of precedence (see \citet[Section 2, pp.865--866]{Kreps_Wilson_1982}).

We let $H_i$ denote the set of \emph{information sets} of player $i$, i.e., this is the partition of all those non-terminal histories where player $i$ is active, where the elements of $H_i$ satisfy the property that, if $x, x' \in h$, with $h \in H_i$, then $A_i (x) = A_i (x')$. We let $H^{\varnothing}_i := H_i \cup \{ \la \varnothing \ra\}$ and $H := \bigcup_{j \in I} H_j$. We extend to information sets the notational convention introduced for histories and we let $A_i (h)$ denote the set of actions available to player $i$ at information set $h \in H_i$.

A \emph{standard strategy} of player $i$ is a function $\mbf{s}_i : H_i \to \bigcup_{\overline{h} \in H_i} A_i (\overline{h})$ such that $\mbf{s}_i (h) \in A_i (h)$, for every $h \in H_i$. We let $\mbf{S}_i$ denote the set of standard strategies of player $i$ and $\mbf{S}_i (h)$ denote the set of standard strategies of player $i$ that reach information set $h \in H$. Thus, given a standard strategy $\mbf{s}_i \in \mbf{S}_i$, we let $H (\mbf{s}_i)$ denote the set of information sets that are not precluded by $\mbf{s}_i$. Given two standard strategies $\mbf{s}_i, \mbf{s}'_i \in \mbf{S}_i$, they are deemed behaviorally equivalent if $H (\mbf{s}_i) = H (\mbf{s}'_i)$ \emph{and} $\mbf{s}_i (h) = \mbf{s}'_i (h)$ for every $h \in H (\mbf{s}_i) = H (\mbf{s}'_i)$. A \emph{strategy}\footnote{\label{foot:strategy}Alternatively called ``plan of actions'' as in the words of  \citet[Section 2]{Rubinstein_1991} or ``reduced strategy''. We use the---nonstandard---expression ``standard strategy'' to refer to an element of $\mbf{S}_i$, for an arbitrary player $i$, where usually such an object is simply called ``strategy'', since we want to save that term for the actual primitive objects of interests for our analysis, which are the elements of $S_i$.} $s_i$ of player $i$  is a maximal set of behaviorally equivalent standard strategies. We let $S_i$ denote the set of all strategies of player $i$ and---following standard notational conventions---we let $S_{-i} := \prod_{j \in I \setminus \{i\}} S_j$, with $S := \prod_{j \in I} S_j$. 

We extend the conventions introduced for standard strategies to strategies, i.e., we let $H (s_i)$ denote the set of information sets compatible with $s_i$, with $H_i (s_i)$ and $H_{-i} (s_i)$ defined accordingly. We let $S_i (h)$ denote the set of strategies of player $i$ that allow information set $h \in H$: given that $S (h) := \prod_{j \in I} S_j (h)$, from assuming perfect recall, we have that $S (h) = S_i (h) \times S_{-i} (h)$, for every $h \in H$. Observe that the notational conventions introduced in the paragraph above extend naturally to $S_{i} (h)$, $S_{-i} (h)$, and $S (h)$. Also, we let $\mcal{H}_i := \Set { S_{-i} (h) | h \in H^{\varnothing}_i}$, for every $i \in I$. Finally, we let $u_i \in \Re^S$ be the \emph{payoff function} of player $i \in I$.

\subsection{Beliefs \& Related Objects}
\label{subsec:beliefs}

In the following, given a compact metrizable space $Y$, we let $\mcal{A} (Y)$ denote its Borel $\sigma$-algebra, with the product $\sigma$-algebra as canonically defined for $Y$ having a product structure and with $\proj$ denoting the projection operator as canonically defined, and we let $\mcal{C} \subseteq \mcal{A} (Y) \setminus \{ \emptyset \}$ denote a countable set of (open and closed) \emph{conditioning events}: the objects just introduced give rise to the \emph{conditional measurable space} $(Y, \mcal{A} (Y), \mcal{C})$. Also, we let $\Delta (Y)$ denote the set of all $\sigma$-additive probability measures (henceforth, probability measures) over an arbitrary space $Y$, with $\marg$ denoting the---continuous---marginal operator  as canonically defined, $\supp \mu$ denoting the support of an arbitrary probability measure $\mu \in \Delta (Y)$, and $\delta_{\{y\}}$ denoting the Dirac measure for an arbitrary element $y \in Y$. Furthermore, we let $\wp (Y)$ denote the power set of an arbitrary set $Y$. 

Given a \emph{conditional measurable space} $(Y, \mcal{A} (Y), \mcal{C})$, a \emph{conditional probability system}\footnote{\label{foot:CPS}This  corresponds to one of the primitive elements of a  \emph{conditional probability space} of \citet[Sections 1.2 \& 1.4]{Renyi_1955}. However, the name---that eventually stuck in the game-theoretic literature---actually comes from a related definition from \citet[Section 5, pp.336--337]{Myerson_1986}. See \citet[Section 3.2]{Hammond_1994} for a discussion of the relation between these two notions.} (henceforth, CPS) $\nu$ on the conditional measurable space $(Y, \mcal{A} (Y), \mcal{C})$ is a mapping
\begin{equation*}
	\nu \Rounds { \cdot | \cdot } : \mcal{A} (Y) \times \mcal{C} \to [0,1]
\end{equation*}
that satisfies the following axioms:
\begin{enumerate}[label=A\arabic*., leftmargin=*, itemsep=0.5ex]
	\item for every $C \in \mcal{C}$, $\nu \Rounds {C | C} =1$;
	\item for every $C \in \mcal{C}$, $\nu \Rounds { \cdot | C } \in \Delta (Y)$;
	\item (Chain Rule) for every $A \in \mcal{A} (Y)$ and for every $B, C \in \mcal{C}$, if $A \subseteq B \subseteq C $, then $\nu (A|C) = \nu (A | B) \cdot \nu (B|C)$. 
\end{enumerate}
We let $\Delta^{\mcal{C}} (Y)$ denote the set of CPSs on $(Y, \mcal{A} (Y), \mcal{C})$.

Following the Bayesian approach to game theory as in \cite{Aumann_1987}, in this paper, probability measures and related objects (such as CPSs) are used to represent players' beliefs. Thus,  given an arbitrary dynamic game $\Gamma$, for every $i \in I$, for our purposes, the relevant conditional measurable space for player $i \in I$ is given by $Y := S_{-i}$, $\mcal{A} (Y) := \wp (S_{-i})$, and $\mcal{C} :=\mcal{H}_i$, where---following the notation introduced above---we let $\Delta^{\mcal{H}_i} (S_{-i})$ denote the set of CPSs on $(S_{-i}, \wp (S_{-i}), \mcal{H}_i)$, with typical element $\mu_i (\cdot | S_{-i} (h))$, for every $h \in H^{\varnothing}_i$.

We say that $\mu_i \Rounds { \cdot | S_{-i} (h)}$ \emph{conditionally believes} event $E_{-i} \subseteq S_{-i}$ if $\mu_i \Rounds { E_{-i} | S_{-i} (h)} = 1$. Thus, we write that $\mu_i \in \Delta^{\mcal{H}_i} (S_{-i})$ \emph{strongly believes} event $E_{-i} \subseteq S_{-i}$ if $\mu_i \Rounds {E_{-i} | S_{-i} (h) } = 1$, for every $h \in H_i$ such that $S_{-i} (h) \cap E_{-i} \neq \emptyset$. As an immediate consequence of the monotonicity of probability measures, the notion of conditional belief satisfies \emph{monotonicity}, i.e., the property that states that for every $E_{-i} \subseteq F_{-i}$ and $S_{-i} (h) \in \mcal{H}_i$, $\mu_i \Rounds { E_{-i} | S_{-i} (h) } \leq \mu_i \Rounds { F_{-i} | S_{-i} (h)}$. On the contrary, \cite{Battigalli_Siniscalchi_2002} show that---interestingly---strong belief does \emph{not} satisfy monotonicity. We illustrate this failure by means of the leading example represented in \Sref{fig:CIC}.

\begin{example}[label=ex:CIC, name=Running Example]
Consider the dynamic game in \Sref{fig:CIC}. Focusing on Bob, we let $\mu_b := (\mu_b \Rounds { \cdot | S_a (\la \varnothing \ra) } , \mu_b \Rounds { \cdot | S_a (\la In \ra) } )$ be defined as follows: $\mu_b \Rounds { \cdot | S_a (\la \varnothing \ra) } := \delta_{\{Out\}}$ and $\mu_b \Rounds { \cdot | S_a (\la In \ra) } := \delta_{\{In\mhyphen Down\}}$. Now, let $E_{a} = \{ Out \}$ and let $F_a := \{ Out, In\mhyphen Across \}$. We have that $E_a \subseteq F_a$, but, whereas  $\mu_b$ strongly believes $E_a$, $\mu_b$ does \emph{not} strongly believes $F_a$. Indeed, $\mu_b$ strongly believes $E_a$ since there is only one information set in $H_b$ such that  $S_{a} (h) \cap E_{a} \neq \emptyset$ , namely, $\la \varnothing \ra \in H_b$, and it is the case that    $\mu_b \Rounds {E_a | S_{a} (\la \varnothing \ra) } = 1$. However, $\mu_b$ does not strongly believe $F_a$, since there exists an information set $h^{*} \in H_b$ such that $F_a \cap S_a ( h^* ) \neq \emptyset$ and $\mu_b \Rounds {F_a | S_{a} (h^*) } \neq 1$, namely, $h^* := \la In \ra$.
\end{example}

Building on the notion of CPS, we say that a strategy $s^{*}_i \in S_i$ is a \emph{sequential best-reply} to CPS $\mu_i \in \Delta^{\mcal{H}_i} (S_{-i})$ if 
\begin{equation*}
\sum_{s_{-i} \in S_{-i}} u_i (s^{*}_i, s_{-i}) \cdot \mu_i \Rounds { \{s_{-i}\} | S_{-i} (h)} \geq %
\sum_{s_{-i} \in S_{-i}} u_i (s_i, s_{-i})  \cdot \mu_i \Rounds { \{s_{-i}\} | S_{-i} (h)},
\end{equation*}
for every $h \in H_i (s^{*}_i)$ and every $s_i \in S_i (h)$. In the following, given a player $i \in I$ and a CPS $\mu_i \in \Delta^{\mcal{H}_i} (S_{-i})$, we let $\rho_i (\mu_i)$ denote player $i$'s set of sequential best-replies to CPS $\mu_i$.

\subsection{Solution Concepts}
\label{subsec:solution_concepts}

We now introduce two solution concepts for dynamic games. The first one is the so-called Strong Rationalizability procedure of \citet[Definition 9, p.1042]{Pearce_1984} and \citet[Definition 2, p.46]{Battigalli_1997}, which is a form of Rationalizability for dynamic games that captures forward induction reasoning, i.e., the idea that a player  is going to make deductions based on the opponents' rational behavior in the past.\footnote{See \citet[Section 2.6, p.1013]{Kohlberg_Mertens_1986} and \citet[Section 2.4, p.8]{Kohlberg_1990}. Alternatively, forward induction can be informally captured via the  \emph{Best Rationalization Principle} of \citet[Section 1.1, p.180]{Battigalli_1996}.}

Formally, given a dynamic game $\Gamma$, for every $i \in I$ and $m \in \bN$, we let $\SR^{0}_{i}  := S_i$, and---assuming that $\SR^{m} := \SR^{m}_{i} \times \SR^{m}_{-i}$ have been defined---we let 
\begin{equation*}
\label{eq:p_def}
\SR^{m+1}_{i}  := %
\Set { s^{*}_i \in \SR^{m}_{i} | %
\begin{array}{l}
\exists \mu_i \in \Delta^{\mcal{H}_i} (S_{-i}) : \\%
1.\  s^{*}_i \in \rho_i (\mu_i) , \\
2.\  \forall h \in H_i \ \Big( S_{-i} (h) \cap \SR^{m}_{-i} \neq \emptyset \Lra \mu_i \Rounds{ \SR^{m}_{-i} | S_{-i} (h) } = 1 \Big)
\end{array}
}.\\
\end{equation*}
Thus, we let $\SR^{\infty}_{i}  := \bigcap_{m \geq 0} \SR^{m}_{i}$ denote the set of strategies of player $i$ that survive the Strong Rationalizability procedure, with $\SR^{\infty} := \prod_{j \in I} \SR^{\infty}_{j}$ denoting the \emph{set of strongly rationalizable strategy profiles}.

\begin{example}[continues=ex:CIC, name=Strong Rationalizability]
The game in \Sref{fig:CIC} allows for a clear uncovering of forward induction reasoning. Informally, according to forward induction reasoning,  upon being called into play, Bob should infer that Ann is---rationally---playing strategy $In\mhyphen Across$ in light of the fact that strategy $In\mhyphen Down$ is going to give her the lowest possible payoff. As a result, Bob should best-respond by playing $Go$.  Now, Strong Rationalizability formalizes the steps above.\footnote{Incidentally, here the informal argument also seems reminiscent of the usual backward induction argument. In dynamic games with perfect information without relevant ties, \citet[Theorem 4, p.53]{Battigalli_1997} states that  Strong Rationalizability  and backward induction reasoning (as captured, for example, by Subgame Perfect Equilibrium) are outcome-equivalent, whereas here the outcome is actually the same as the strategy profile.}  Indeed, there exists no CPS for Ann such that $In\mhyphen Down$ is a sequential best-reply for her, leading to $\SR^{1}_a = \{ Out, In\mhyphen Across \}$. As a result, we have $\SR^{2}_b = \{ Go \}$, because there are no CPSs for Bob that make $Stop$ a sequential best-reply given the previous step. Finally, we have $\SR^{3}_a = \{In\mhyphen Across\} = \SR^{\infty}_a$ and $\SR^{3}_b = \{Go\} = \SR^{\infty}_b$, as already informally stated in the introduction.
\end{example}

The second solution concept we introduce is that of Full Strong Best-Reply Sets of \cite{Battigalli_Friedenberg_2012}.

\begin{definition}[Full Strong Best-Reply Sets (FSBRSs)]
\label{def:full_SBRS}
Given a dynamic game $\Gamma$, a product set $F := \prod_{j \in I} F_j \subseteq S$ is a \emph{full strong best-reply set} (henceforth, FSBRS) if for every $i \in I$ and $s^{*}_i \in F_i$ there exists a CPS $\mu_i \in \Delta^{\mcal{H}_i} (S_{-i})$ such that:
\begin{enumerate}[label=\arabic*)]
\item $s^{*}_i \in \rho_i (\mu_i)$;

\item $\mu_i$ strongly believes $F_{-i}$;

\item $s_i \in F_i$, for every $s_i \in \rho_i (\mu_i)$.
\end{enumerate}
\end{definition}

Regarding Conditions (1) and (2), we say that the CPS $\mu_i$ \emph{justifies}  strategy $s^{*}_i$ within $F$. The characterizing condition of \emph{fullness} is Condition (3), which captures a form of ``closure under sequential best-replies'' property.\footnote{A similar closure property appears in the notion of \emph{Closed Under Rational Behavior (CURB) sets} as introduced by \citet[Section 2, p.143]{Basu_Weibull_1991}. Non-full Strong Best-Reply Sets (called ``Extensive-Form Best-Reply Sets'' in \citet[Definition 12.29, p.680]{Dekel_Siniscalchi_2015}) do not play any role in our analysis and are therefore not introduced separately. }

We now introduce the first novel notion of this paper, which, building on \Mref{def:full_SBRS}, focuses on the perspective of a specific player.

\begin{definition}[Player-Specific FSBRSs]
\label{def:player-specific_FSBRS}
Given a dynamic game $\Gamma$ and a player $i \in I$, a set $F_i \subseteq S_i$ is a \emph{player $i$'s specific FSBRS} if there exists an $F_{-i} := \prod_{j \in I \setminus \{i\}} F_j$ such that $F_i \times F_{-i}$ is an FSBRS.
\end{definition} 

In the following, given that we let $\FSBRS$ denote the family of FSBRSs of a dynamic game $\Gamma$, it is immediate to observe that the family of player $i$'s specific FSBRSs, denoted $\FSBRS_i$, can be defined as  
\begin{equation*}
\FSBRS_i := \Set { F_i \in \wp (S_i) | \exists F \in \FSBRS : \proj_i F = F_i },
\end{equation*}
for every $i \in I$. It is crucial to observe that an element of $\prod_{j \in I} \FSBRS_j$ might not be an FSBRS itself: this, and the definitions themselves, are illustrated next via our leading example.

\begin{example}[continues=ex:CIC, name=FSBRSs]
Focusing on the dynamic game in \Sref{fig:CIC},  we have
\begin{equation*}
\FSBRS = \{
		\{ In\mhyphen Across\} \times \{ Go\} , %
		\{ Out \} \times \{Stop \} ,%
		\{ Out \} \times \{Stop,Go \}, %
		\{ \emptyset \}
\},
\end{equation*}
which also gives us
\begin{align*}
	\FSBRS_a & =  \{ %
		\{ In\mhyphen Across \}  , %
		\{ Out \},
		\{ \emptyset \}
	\} , \\
	\FSBRS_b & =  \{ %
	\{ Go \} , %
	\{ Stop \} , %
	\{ Stop, Go \} , %
	\{ \emptyset \}
\}.
\end{align*}

\noindent Thus, regarding $\FSBRS$, whereas the first FSBRS above captures the strategy profile obtained via $\SR^\infty$, it is immediate to observe that there are additional FSBRSs that are unrelated to the strategy profile obtained via $\SR^\infty$. Also, concerning player-specific FSBRSs, we have $\{ In\mhyphen Across \} \times \{ Stop \} \in \FSBRS_a \times \FSBRS_b$, which is \emph{not} an element of $\FSBRS$.
\end{example}

Building on the previous example, we now collect in a dedicated remark (for future reference) the fact that it is possible to have FSBRSs that do \emph{not} coincide with the strategy profiles obtained via Strong Rationalizability.

\begin{remark}[Strong Rationalizability \& FSBRSs]
\label{rem:SR_FSBRS}
It is not necessarily the case that, given a dynamic game $\Gamma$ and its family of FSBRSs $\FSBRS$, every $F \in \FSBRS$ is such that $F \subseteq \SR^\infty$.
\end{remark}

\subsection{Infinite Hierarchies of Beliefs \& Type Structures}
\label{subsec:hierarchies}

Given a dynamic game, as outside observers in the role of analysts, we are interested in analyzing interactive reasoning in such a context. Thus, in order to perform our analysis, first of all we \emph{exogenously} select the players' infinite hierarchies of beliefs that are of interest.

For our purposes, given an arbitrary player $i \in I$, we simply need to recall the definition of CPSs provided in \Sref{subsec:beliefs} on the conditional measurable space $(S_{-i}, \wp (S_{-i}), \mcal{H}_i)$: this conditional measurable space is actually the \emph{first} relevant domain of uncertainty of player $i$ upon which we inductively define all player $i$'s relevant domains of uncertainty---along with corresponding conditioning events---as follows:
\begin{center}
	\begin{tabular}{cc}
		$ X_i^{0} := S_{-i}$, 		&%
		$ \mcal{C}_i^{0} :=  \mcal{H}_i$, \\
		$\vdots$			& 	$\vdots$ \\
		$ X^{m+1}_i = X^{m}_i \times \Delta^{\mcal{C}_i^{m}} (X_{-i}^{m}) $, &%
		$\mcal{C}_i^{m+1} = \Set {  D \subseteq X_i^{m+1} | \exists C \in \mcal{C}_i^{m} : D = C \times \Delta^{\mcal{C}_i^{m}} (X^{m}_{-i}) } $,   \\
		$\vdots$		&	 	$\vdots$ \\
	\end{tabular}
\end{center}
Now, for every $m \geq 1$ and every subset $A   \in \mcal{C}$, let $\mcal{D}^{m} (A) := A \times \prod^{m - 1 }_{\ell = 0} \Delta^{\mcal{C}^m_i} (X^{\ell}_{-i})$
and define an \emph{infinite hierarchy of coherent beliefs} (henceforth, IHB) of player $i$ as a sequence $\nu_i := (\nu_i^{n})_{n \in \bN}$, with $\nu^{k}_i \in \Delta^{\mcal{C}^{k-1}_i} (X_{-i}^{k-1})$ such that
\begin{equation*}
	\marg_{X^{k-1}} \nu^{k+1}_i \Rounds { \cdot | \mcal{D}^k (C) } = %
	\nu^{k}_i \Rounds { \cdot | \mcal{D}^{k-1} (C) } ,
\end{equation*}
for every $C \in \mcal{C}^{0}_i$ and $k \in \bN$. In particular, as it is widespread in the literature, we focus on those IHBs that satisfy the additional condition of \emph{common belief in coherency}:\footnote{See \cite{Battigalli_et_al_2020} for an exception that studies possibly incoherent infinite hierarchies of beliefs.} i.e., it is understood that those IHBs we refer to in this work satisfy this additional condition. In general, the interested reader is referred to \citet[Section 2]{Battigalli_Siniscalchi_1999} for the details of the construction and, in particular, to Section 2.4 therein for the formal definition of common belief in coherency.\footnote{It has to be observed that \cite{Battigalli_Siniscalchi_1999} use the word ``certainty'' where we use the word ``belief''.}

Instead of handling IHBs directly, since the seminal \cite{Harsanyi_1967-1968}, it is common to use types and type structures to describe IHBs in a `compact' way. Thus, given a dynamic game  $\Gamma$, a \emph{standard epistemic type structure with conditioning events} (henceforth, standard type structure---see \cite{Ben-Porath_1997} and \cite{Battigalli_Siniscalchi_1999}) appended on $\Gamma$ is a tuple
\begin{equation*}
	\mscr{T} := \la  (T_{j} , \beta_{j} )_{j \in I} \ra,
\end{equation*}
where, for every $i \in I$, 
\begin{itemize}[leftmargin=*]
	\item $T_i$ is her compact metrizable space of \emph{epistemic types} (henceforth, types) and
	
	\item  $\beta_i := (\beta_{i, h})_{h \in H^{\varnothing}_i} : T_i \to  \Delta^{\mcal{H}_i} (S_{-i} \times T_{-i})$ is her continuous \emph{belief function}. 
\end{itemize}
A standard type structure is \emph{finite} if $T_i$ is finite, for every $i \in I$.

\begin{example}[continues=ex:CIC, name=A Finite Standard Type Structure]
We focus again on the dynamic game in \Sref{fig:CIC} and we append on it a (finite) standard type structure $\mscr{T}$. Let $T_i := \{ t_i , t'_i \}$, with $i \in \{ a , b \}$, be the type spaces of Ann (viz., $a$) and Bob (viz., $b$). The players' belief functions, which capture the players' beliefs at all relevant information sets, are described in the tables below.

\begin{table}[H]
\hfill
\parbox{.45\linewidth}{
\centering
\begin{tabular}{@{}rcc@{}} \toprule
	& $\beta_{a, \la \varnothing \ra} (\overline{t}_a)$ & $\beta_{a, \la In, Go \ra} (\overline{t}_a)$ \\ 
	\midrule
	$(In\mhyphen Across, t_a)$ & $(0, 0, 1)$ & $(0, 0, 1)$ \\  
	$(Out, t'_a)$ & $(1, 0, 0)$ & $(0,1, 0)$ \\ 
	$(In\mhyphen Down, t'_a)$ & $(1, 0, 0)$ & $(0,1, 0)$ \\  
	\bottomrule
\end{tabular}
}
\hfill
\parbox{.45\linewidth}{
\centering
\begin{tabular}{@{}rcc@{}} \toprule
	& $\beta_{b, \la \varnothing \ra} (\overline{t}_b)$ & $\beta_{b, \la In \ra} (\overline{t}_b)$ \\ 
	\midrule
	$(Stop, t_b)$ & $(0, 1, 0)$ & $(0, 0, 1)$ \\
	$(Go, t_b)$ & $(0, 1, 0)$ & $(0, 0, 1)$ \\
	$(Go, t'_b)$ & $(1, 0, 0)$ & $(1, 0, 0)$ \\     
	\bottomrule
\end{tabular}
}
\label{tab:belief_function_01}
\caption{Types' beliefs captured via players' belief functions in the standard type structure $\mscr{T}$.}
\end{table}

\noindent To illustrate these tables, for example, consider Ann's type $t'_a$: at  $\la \varnothing \ra$, this type believes  that Bob is of type $t_b$ and he is going to play $Stop$; however, when faced with her second decision (i.e., at $\la In \ra$), whereas it still believes that Bob's type is $t_b$, \emph{a fortiori} it changes its belief concerning the play by believing $(Go, t_b)$. Focusing on Bob, for example Bob's type $t'_b$ believes, both at $\la \varnothing \ra$ and $\la I \ra$, that Ann is of type $t_a$ and she is going to play $In\mhyphen Across$.
\end{example}

For every dynamic game $\Gamma$, we let  $\mscr{T}^* := \la ( T^*_{j} , \beta^{*}_{j} )_{j \in I} \ra$ denote the \emph{canonical hierarchical structure} appended on $\Gamma$, constructed as in \citet[Section 2]{Battigalli_Siniscalchi_1999} as the space that  comprises all the players' IHBs as sketched above.\footnote{See \citet[Section 2]{Brandenburger_Dekel_1993} for the construction without conditioning events.} As shown in \citet[Proposition 3, p.202]{Battigalli_Siniscalchi_1999}, the canonical hierarchical structure $\mscr{T}^*$ is a standard type structure in its own rights that is \emph{universal} according to the terminology introduced by \cite{Siniscalchi_2008}, that is:
\begin{itemize}[leftmargin=*]
\item it is  \emph{terminal}, since every other standard type structure can be uniquely embedded in it, and 

\item \emph{belief-complete}, since the belief function $\beta^{*}_i$ is surjective, for every $i \in I$.\footnote{This completeness notion was introduced by \cite{Brandenburger_2003}.}
\end{itemize}
As a result, we call $\mscr{T}^*$ the \emph{universal type structure}.

As $\mscr{T}^*$ is the universal type structure, there is a unique type-morphism\footnote{For the definition of type-morphism see, for example, \citet[Definition 3, p.201]{Battigalli_Siniscalchi_1999}.} from any standard type structure $\mscr{T}$ into $\mscr{T}^*$, which maps a type in $\mscr{T}$ to its corresponding IHB, which is an element of $\mscr{T}^*$.\footnote{Again, details can be found in \citet[Proposition 3, p.202]{Battigalli_Siniscalchi_1999}.} As a matter of fact, here this correspondence is injective, because we only consider non-redundant type structures (i.e., redundant types do not play any role in our analysis); also, the universal type structure is non-redundant by construction. Thus, we can uniquely identify types with their corresponding IHBs. Since our work focuses only on IHBs, we employ the convention that we always consider the type space where each type itself is just the associated IHB. This allow us to save on notation without introducing (and using) type-morphisms explicitly. However, in concrete examples, we refrain from always specifying the full IHB, but work instead with the usual `type as arbitrary label' convention: it should be understood that these labels are a representation of the IHBs themselves.\footnote{This is in line with the interpretation of types as in \citet[Sections 12.1.1, 12.2.3, \& 12.2.4]{Dekel_Siniscalchi_2015}, to which the interested reader is referred for further methodological discussions about this point.}

\subsection{Interactive Epistemology with Standard Type Structures}
\label{subsec:interactive_epistemology}

To capture \emph{dynamic} interactive reasoning in standard type structures, we now introduce modal operators. For this purpose, we introduce the---usual--- definitions in the context of standard type structures, with the understanding that those very same operators' definitions are used in the analysis we perform in what follows, which is based on a a generalization of the present framework. Thus, for this part, consider a standard type structure $\mscr{T}=\la  (T_{j} , \beta_{j} )_{j \in I} \ra$ as fixed, which itself is appended to a fixed dynamic game $\Gamma$. Hence, for every $i \in I$, $h \in H^{\varnothing}_i$, and $E_{-i} \in \mcal{A} (S_{-i}\times T_{-i})$, the \emph{conditional belief operator} $\Bel_{i, h}$ of player $i$ is defined as 
\begin{equation*}
\Bel_{i, h} (E_{-i}) := \Set { (s_i , t_i) \in S_i \times T_i |  \beta_{i, h} (t_i) (E_{-i}) = 1  },
\end{equation*}
while the \emph{strong belief operator} $\SB_i$ of player $i$ is defined as
\begin{equation}
\label{eq:strong_belief}
\SB_i (E_{-i}) := \bigcap_{h \in H^{\varnothing}_i : \ E_{-i} \cap [S_{-i} (h) \times T_{-i}] \neq \emptyset} \Bel_{i, h} (E_{-i}),
\end{equation}
with $\SB_i (\emptyset) = \emptyset$ and $\SB (E) := \prod_{i \in I} \SB_{i} (E_{-i})$, for every $E \in \prod_{j \in I} \mcal{A} (S_j \times T_j)$, where it has to be observed that the conditional belief operator and the strong belief operator inherit their relation to the monotonicity property from the corresponding belief notions as in \Sref{subsec:beliefs}. However, whereas the (conditional) belief operator essentially only depends on a type (i.e., the IHB) itself, the strong belief operator does also depend on the `ambient' type structure (i.e., all the types, along with their beliefs, that are elements of a given type structure of interest). That is, the \emph{context} of the game matters and this has crucial implications for the analysis that we perform in this work.\footnote{Equivalently, one could start just with IHBs (again) with the idea of looking for strong belief not just in an event $E$, but also in conjunction with a---formally transparent---restriction on what players actually reason about. See \cite{Battigalli_Friedenberg_2012} and \cite{Battigalli_Prestipino_2013} for details.\label{footnote:transparent_restriction}}

Finally, we define the \emph{common correct strong belief} operator $\CSB (E) := E \cap \SB (E)$,\footnote{Note that, as usual in models without introspective beliefs, we do impose correct beliefs to restrict behavior in addition to restricting (higher-order) beliefs. The arguments made in the first paragraph in \citet[Section 13.3.2]{Dekel_Siniscalchi_2015} apply verbatim to our framework. This is not to be confused with imposing the Truth Axiom, which would impose correct beliefs for all possible events.} whose iterated application---obviously---abides by the rules that govern the iteration of the application of an arbitrary operator $\mbb{O}$ on an arbitrary event $E$, i.e.,  $\mbb{O}^0 (E) := E$ and $\mbb{O}^n (E) := \mbb{O} ( \mbb{O}^{n-1} (E) )$, for every $n \in \bN$. In what follows, we do not employ the word ``correct'' unless explicitly needed.

With modal operators at our disposal, we now define the event of our interest, namely, that of \emph{rationality of a player} $i \in I$, which builds on the notion of sequential best-reply to a CPS:
\begin{equation*}
\label{eq:rat_standard}
\Rat_i := \Set { (s^{*}_i , t_i) \in S_i \times T_i |   s^{*}_i \in \rho_i (\marg_{S_{-i}}\beta_i (t_i)) },
\end{equation*}
with $\Rat :=\prod_{j \in I} \Rat_j$. This leads straightforwardly to our epistemic event of interest: given that, for $m \in \bN$, we define $\CSB^{m} (\Rat)  :=  \Rat \cap \bigcap^{m - 1}_{k = 0} \SB (\CSB^{k} (\Rat))$ as the event that captures \emph{Rationality and $m$-order Strong Belief in Rationality}, we let
\begin{equation*}
\label{eq:RCSBR}
\RCSBR  :=  \CSB^{\infty} (\Rat)  :=  \bigcap_{\ell \geq 0} \CSB^\ell (\Rat) 
\end{equation*}
denote the event capturing \emph{Rationality and Common Strong Belief in Rationality} (henceforth, RCSBR).

Given the framework introduced above, we can recall a crucial game-theoretical result that is the basis for our work, whose importance lies in connecting RCSBR and FSBRSs. Indeed, the following result actually establishes that FSBRSs characterize RCSBR in arbitrary standard type structures.

\begin{theorem}[\cite{Battigalli_Friedenberg_2012}]
\label{rem:characterization}
Given a dynamic game $\Gamma$, the following statements hold:\footnote{Whereas their result is stated for $2$-player games only, the authors point out in \citet[Section 9.c]{Battigalli_Friedenberg_2012} that their \citet[Theorem 1, pp.66--67]{Battigalli_Friedenberg_2012} holds for dynamic games with three or more players up to issues of correlation. In our framework this correlation issue is void.} 
\begin{enumerate}[label=\arabic*)]
\item for every standard type structure $\mscr{T}$, $\proj_S \RCSBR \in \FSBRS$;

\item for every $F \in \FSBRS$, there exists a finite standard type structure $\mscr{T}$  such that $\proj_S \RCSBR = F$.
\end{enumerate}
\end{theorem}

Two points should be emphasized regarding \Mref{rem:characterization}. First of all, a crucial element of \emph{Full} SBRSs is the `closure' Condition (3) in \Mref{def:full_SBRS}, with  \citet[Example 7, pp.65--66]{Battigalli_Friedenberg_2012} showing this in detail. Also, it should be observed that \citet[Proposition 6, p.373]{Battigalli_Siniscalchi_2002} shows how $\proj_S \RCSBR = \SR^\infty$ in $\mscr{T}^*$, which implies that the set of Strongly Rationalizable strategy profiles is always an FSBRS.

\subsection{State Spaces}
\label{subsec:states}

Recalling that $\mscr{T}^*$ denotes the universal type structure, it is customary to let $\Omega^* := \prod_{j \in I} S_j \times \prod_{j \in I} T^{*}_j$ denote the \emph{universal state space}. More generally, we have the following definition of a state space.

\begin{definition}[State Space]
\label{def:state_space}
A \emph{state space} is a measurable subset $\widetilde{\Omega}  := \prod_{j \in I} S_j \times \prod_{j \in I} \widetilde{T}_j  \subseteq \Omega^{*}$ of the universal state space.
\end{definition}

The following notion is a translation in the framework of standard type structures of \citet[Definition 2.15, p.12]{Mertens_Zamir_1985}: a state space $\widetilde{\Omega}$  is \emph{belief-closed} if
\begin{equation}
\label{eq:belief-closed}
\supp \beta^{*}_{i,h} (t_i) \subseteq S_{-i} \times \widetilde{T}_{-i},
\end{equation}
for every $i \in I$, $h \in H^{\varnothing}_i$, and $t_i \in \widetilde{T}_i$. Note that this definition does not only ask for belief-closedness at the initial history, but also at every history for a player. Recalling our convention on types in standard type structures as representations of IHBs, we can naturally define a state space from a standard type structure $\mscr{T} := \la  (T_{j} , \beta_{j} )_{j \in I} \ra$ with $\prod_{j \in I} S_j \times \prod_{j \in I} T_j$. The resulting state space is belief-closed as shown---in a slightly different setting---in \citet[Remark 2, p.201]{Battigalli_Siniscalchi_1999}. The converse holds, i.e., that a belief-closed state space induces a standard type structure is again immediate.\footnote{Exactly as it is immediate the analogous result stated in \citet[Proposition 2.16, p.13]{Mertens_Zamir_1985}.} For latter reference, we state this formally.

\begin{remark}
\label{remark:structures_closed_spaces}
The following hold.
\begin{enumerate}[label=\arabic*)]
\item For every standard type structure $\widetilde{\mscr{T}} := \la  (\widetilde{T}_{j} , \beta_{j} )_{j \in I} \ra$ there exists a belief-closed state space: $\widetilde{\Omega}  := \prod_{j \in I} S_j \times \prod_{j \in I} \widetilde{T}_j$.

\item For every belief-closed state space $\widetilde{\Omega}  := \prod_{j \in I} S_j \times \prod_{j \in I} \widetilde{T}_j$ there exists a standard type structure $\widetilde{\mscr{T}} := \la  (\widetilde{T}_{j} , \beta_{j} )_{i \in I} \ra$ such that $\beta_i (t_i) = \beta_i^* (t_i)$, for every $i \in I$ and $t_i \in \widetilde{T}_i$. 
\end{enumerate}
\end{remark}

We can now formalize the---central in this paper---notion of non-belief-closed state space.

\begin{definition}[Non-Belief-Closed State Space]
\label{def:non-belief-closed}
A state space $\widetilde{\Omega} :=  \prod_{j \in I} S_j \times \prod_{j \in I} \widetilde{T}_j$ is \emph{non-belief-closed} if there exists a player $i \in I$, an information set $h \in H^{\varnothing}_i$, and a type $t_i \in \widetilde{T}_i$ such that 
\begin{equation*}
\proj_{T_{-i}} \supp \beta^{*}_{i,h} (t_i) \not\subseteq \widetilde{T}_{-i}.
\end{equation*}
\end{definition}

Again, it has to be observed that this definition allows for belief-closedness at some histories, but the state space is deemed non-belief-closed if at least for one player there is one information set where she `reasons' outside the given state space.

\begin{example}[continues=ex:CIC, name=Non-Belief-Closedness]
Recalling the finite standard type structure we appended previously on the dynamic game in \Sref{fig:CIC}, it is immediate to see that the---naturally defined, as sketched above---state space $S_a \times T_a \times S_b \times T_b$ with beliefs as described in \hyperref[tab:belief_function_01]{\color{burgundy}{Table 1}}  satisfies \Mref{eq:belief-closed}. Indeed, this is immediate by the argument above and, in particular, by \citet[Remark 2, p.201]{Battigalli_Siniscalchi_1999}.  On the contrary, if we now consider $\Omega := S \times \{t_a\} \times \{t_b\}$  as the state space of interest, it is immediate to observe that this state space is \emph{not} belief-closed, i.e., it does not satisfy \Mref{eq:belief-closed}.
\end{example}

Two observations are in order concerning \Mref{def:state_space} that are crucial for our approach. The notion of state space is
\begin{enumerate}[label=\arabic*)]
 \item based on working with the universal state space $\Omega^*$ and
 
 \item it is \emph{independent} of the notion of type structure. 
\end{enumerate}
Point (1) above is particularly important, since it tells us that we always work with IHBs, which is the actual primitive notion of our endeavour. Point (2) is somewhat at odds with the standard way of introducing the notion of state space in the context of \emph{product} standard type structures (which is the present one), since---typically---the notion of state space is \emph{derived} from all the types of a given standard type structure (in the spirit of Point (1) of \Mref{remark:structures_closed_spaces}). As a matter of fact, the reason for this---apparent---oddity lies exactly in \Mref{def:non-belief-closed}. Indeed, interestingly, the notion of non-belief-closed state space formalized in \Mref{def:non-belief-closed} has never been the focus of any work in the literature on epistemic game theory. Thus, if we take IHBs (as we do here) as the primitive notion of an epistemic game-theoretical endeavour, besides needing the  somewhat standard reverse process, a closure of non-belief-closed state spaces is necessary. These steps can be summarized as: first, the IHBs of interests are selected; then a belief-closed state space is obtained from those IHBs; finally, the corresponding type structure is derived. However, as noted before, for an analysis based on the notion of \emph{strong belief}, typically leading to capturing the behavioral implications of $\RCSBR$, the ambient type structure---the context---crucially matters, since it determines what players are `allowed' to reason about (even beyond the usual assignment of zero probability).\footnote{See also \Sref{footnote:transparent_restriction}.} Since the ambient type structure is not given by the IHBs itself, some care has to be taken about what type structures are considered in this reverse process. We discuss and formalize these issues and corresponding derived objects from our primitive notions of IHBs and state spaces next.

\section{Derived Objects}
\label{sec:derived_objects}

\subsection{Separating Type Structures}
\label{subsec:separating_type_structures}

Imagine to fix a state space $\widetilde{\Omega} : = \prod_{j \in I} S_j \times \prod_{j \in I} \widetilde{T}_j \subseteq \Omega^{*}$, which leads, for every $i \in I$, to a collection of IHBs. By working with standard type structures, we---implicitly---abide by the fact that the objects we employ in our analysis are belief-closed as mentioned above. As a result, we are implicitly assuming as modelers that players in a sense cannot be wrong with respect to their beliefs.

\begin{example}[continues=ex:CIC, name=Non-Belief-Closedness]
Recall again the finite standard type structure we appended on the dynamic game in \Sref{fig:CIC}. Now, imagine that, as outside observers, we know that the actual state space of interest is $\widetilde{\Omega} := S \times \{ t_a \} \times \{t_b\}$ with beliefs described in \hyperref[tab:belief_function_02]{\color{burgundy}{Table 2}} below, which is exactly like \hyperref[tab:belief_function_01]{\color{burgundy}{Table 1}} with the difference that we now emphasize the fact that we are working with IHBs belonging to the universal type structure $\mscr{T}^*$.

 \begin{table}[H]
\hfill
\parbox{.45\linewidth}{
\centering
\begin{tabular}{@{}rcc@{}} \toprule
	& $\beta^{*}_{a, \la \varnothing \ra} (\overline{t}_a)$ & $\beta^{*}_{a, \la In, Go \ra} (\overline{t}_a)$ \\ 
	\midrule
	$(In\mhyphen Across, t_a)$ & $(0, 0, 1)$ & $(0, 0, 1)$ \\  
	$(Out, t'_a)$ & $(1, 0, 0)$ & $(0,1, 0)$ \\ 
	$(In\mhyphen Down, t'_a)$ & $(1, 0, 0)$ & $(0,1, 0)$ \\  
	\bottomrule
\end{tabular}
}
\hfill
\parbox{.45\linewidth}{
\centering
\begin{tabular}{@{}rcc@{}} \toprule
	& $\beta^{*}_{b, \la \varnothing \ra} (\overline{t}_b)$ & $\beta^{*}_{b, \la In \ra} (\overline{t}_b)$ \\ 
	\midrule
	$(Stop, t_b)$ & $(0, 1, 0)$ & $(0, 0, 1)$ \\
	$(Go, t_b)$ & $(0, 1, 0)$ & $(0, 0, 1)$ \\
	$(Go, t'_b)$ & $(1, 0, 0)$ & $(1, 0, 0)$ \\     
	\bottomrule
\end{tabular}
}
\label{tab:belief_function_02}
\caption{Players' IHBs in state space $\widetilde{\Omega}$.}
\end{table}
  
\noindent Now,  Ann's \emph{only} IHB is captured by type $t_a$. However, we also know that there is \emph{no} IHB of Bob that corresponds to type $t'_b$: i.e., we know that Bob is not going to \emph{actually} have this IHB. This clearly leads to a lack of belief-closedness. Rather, Bob's \emph{only} IHB is captured by type $t_b$, which again leads to lack of belief-closedness, since we know---by assumption---that type $t'_a$, which is contemplated by type $t_b$, does not capture any IHB of Ann.
\end{example}

What we want to have is a framework which permits us to capture handily IHBs, at the same time allowing us to distinguish between those that are actual IHBs as originally contemplated by us---as analysts---from those that are simply a byproduct of other IHBs. As a first step, one might think that the following construction would provide such a framework. Given a state space $\widetilde{\Omega}$, a `closure' of $\widetilde{\Omega}$, could be a belief-closed state space $\widehat{\Omega} := \prod_{j \in I} S_j \times \widehat{T}_j$ such that, for every $i \in I$, $\widetilde{T}_i \subseteq \widehat{T}_i$, with the understanding that only the types in $\widetilde{T}_i$ are considered for deriving the behavioral implications of suitable epistemic events. However, even if such a `closure' would be chosen in a minimal way, it would always force each player to reason about $\widetilde{T}_{-i}$, i.e., all the opponents' types \emph{considered by the analyst}. However, it might be the case that a player does not actually reason about some types in $\widetilde{T}_{-i}$. The next example illustrates this point.

\begin{example}[continues=ex:CIC, name=Forced Reasoning]
We focus again on the dynamic game in \Sref{fig:CIC} with state space $\widetilde{\Omega} := S \times \{ t_a \} \times \{t_b\}$ and beliefs described as in \hyperref[tab:belief_function_02]{\color{burgundy}{Table 2}}. Now, a `closure' could actually be the state space with $T_i := \{ t_i , t'_i \}$ for every $i \in I$. Within this (belief-closed) state space, we have $\Rat_a = \{(In\mhyphen Across, t_a), (Out, t'_a)\}$ and $\Rat_b = \{(Stop, t_b) , (Go, t'_b) \}$. And then, $\RCSBR =\CSB^{1} (\Rat) = \{(In\mhyphen Across, t_a,Go, t'_b)\}$. Note, in particular, that to obtain $\CSB^{1} (\Rat)$ we have to dismiss $(Stop, t_b)$, because within this `closure' $t_b$ could believe  $(In\mhyphen Across, t_a)$, but it does not. Thus, looking at the types under consideration of the underlying state space we started out with, we---as analysts---would conclude that $In\mhyphen Across$ is the behavioral implication of RCSBR for Ann, whereas for Bob this implication is empty. However, starting again by looking at Bob's state $(Stop, t_b)$, there is nothing in the IHB corresponding to $t_b$ that tells us that he \emph{needs} to consider $(In\mhyphen Across, t_a)$. Nonetheless, our construction of the `closure' \emph{forces} him to consider this type (and corresponding state). Indeed, from Bob's perspective as given by $t_b$, it might very well be that he does not---under any circumstance---consider $(In\mhyphen Across, t_a)$. For example, from his perspective the relevant closure might be given just by $S \times \{t'_a\} \times \{t_b\}$, in which case the behavioral implications of RCSBR for Bob's type under consideration would be $Stop$.
\end{example}

To allow for the possibility of players reasoning in a less comprehensive way than us as analysts (given what we consider possible), we need a definition of closure that could depend on each individual player. The definition that comes next achieves exactly this objective.

\begin{definition}[Player-Specific Closure of a State Space]
\label{def:closure}
Given a state space $\widetilde{\Omega}  := \prod_{j \in I} S_j \times \prod_{j \in I} \widetilde{T}_j$ and a player $i \in I$, a \emph{closure of $\widetilde{\Omega}$ for player $i$}, denoted $\cl_i(\widetilde{\Omega})$, is a belief-closed state space $\widetilde{\Upomega}^i := \prod_{j \in I} S_j \times \tens{T}^i_j$ such that $\widetilde{T}_i \subseteq \tens{T}^i_i$, where:
\begin{itemize}[leftmargin=*]
\item $\widetilde{T}_i$ is the set of \emph{real} types for player $i$,

\item  $\tens{T}^{c}_i:=\tens{T}^i_i\setminus \widetilde{T}_i$ is her set of \emph{imaginary} types.
\end{itemize}
A player-specific closure $\cl_i (\widetilde{\Omega})$ is \emph{degenerate} if $\tens{T}^{c}_{i} = \emptyset$ and is \emph{non-degenerate} otherwise. 
\end{definition}

\begin{remark}[Expressibility, Existence, and Uniqueness]
\label{rem:existence_uniqueness}
Given a state space $\widetilde{\Omega}$ and a player $i \in I$, the following hold:
\begin{enumerate}[label=\arabic*)]
\item $\cl_i(\widetilde{\Omega})$ exists;
\item $\cl_i(\widetilde{\Omega})$ is compact (\emph{a fortiori} measurable);
\item $\cl_i(\widetilde{\Omega})$ is not necessarily unique.
\end{enumerate}
\end{remark}

Regarding Point (1) in \Mref{rem:existence_uniqueness}, for example, the universal state space is one such a separating closure by appropriately classifying the IHBs in those that correspond to real types and those that correspond to the imaginary ones. Concerning Point (2), compactness follows from \Mref{remark:structures_closed_spaces}. Finally, with respect to Point (3), which concerns the---potential---multiplicity of separating closures for a given state space, for example, if the state space is belief-closed, then it would be its own separating closure with $\tens{T}^{c}_i=\emptyset$.

It is now most natural to provide a definition of type structure stemming from \Mref{def:closure} that `respects' the distinction between real and imaginary types. The next definition achieves exactly this point,

\begin{definition}[Separating Type Structure]
\label{def:separating_type_structure}
Fix a state space $\widetilde{\Omega}$, a player $i \in I$, and player $i$'s separating closure $\cl_i(\widetilde{\Omega})$. The \emph{separating type structure induced from the closure $\cl_i(\widetilde{\Omega})$} is the tuple
\begin{equation*}
\mfrak{T} (\cl_i(\widetilde{\Omega})) := \la (\tens{T}^i_{j} , \bm{\beta}^i_{j} )_{j \in I} \ra ,
\end{equation*}
where $\bm{\beta}^i_j (\tens{t}_j) := \beta^{*}_j (\tens{t}_j)$, for every  $\tens{t}_j \in \tens{T}_j^i$ and $j \in I$. A separating type structure induced from the closure $\cl_i(\widetilde{\Omega})$ is \emph{degenerate} whenever $\cl_i(\widetilde{\Omega})$ is degenerate.
\end{definition}

When the closure $\cl_i (\widetilde{\Omega})$ is clear from the context, since it should not lead to any confusion, we write $\mfrak{T}_i$ instead of $\mfrak{T} (\cl_i(\widetilde{\Omega}))$ and call it a \emph{separating type structure for player $i$} (induced by $\widetilde{\Omega}$) or the \emph{corresponding (to $\widetilde{\Omega}$) separating type structure for player $i$}, where it should be observed that $\mfrak{T}_i$ is a well-defined standard type structure by \Mref{remark:structures_closed_spaces}, because $\cl_i(\widetilde{\Omega})$ is belief-closed. Given this notional convention, we introduce a notion of `agreement' about the type structures within which each player reasons.

\begin{definition}[Common Separating Type Structure]
\label{def:common_type_structure}
Given a state space $\widetilde{\Omega}$, a profile of corresponding separating type structures $(\mfrak{T}_j)_{j \in I}$ is a \emph{common separating type structure}, denoted $\mfrak{T}$, if, for every $i, j, k \in I$ and separating type structures $\mfrak{T}_i$ and $\mfrak{T}_j$, 
\begin{itemize}[leftmargin=*]
\item $\tens{T}^{i}_k = \tens{T}^{j}_k$,

\item $\bm{\beta}^{i}_k (\tens{t}_k) = \bm{\beta}^{j}_k (\tens{t}'_k)$, for every $\tens{t}_k \in \tens{T}^i_k$ and $\tens{t}'_k \in \tens{T}^j_k$ such that $\tens{t}_k = \tens{t}'_k$.
\end{itemize}
Otherwise, the separating type structures are \emph{non-common}.
\end{definition}

We now introduce a notational convention that should make particularly perspicuous the identification of real and imaginary types in specific examples.

\begin{notation}[Bookkeeping within a Closure]
\label{not:union}
Given $i \in I$ and a closure $\cl_i(\widetilde{\Omega})$, we let $\tens{T}^{\heart}_i  :=  \widetilde{T}_i $ and $\tens{T}^{\spadesuit}_{i} :=  \tens{T}^{c}_i$, where we typically write $\tens{t}_i \in \tens{T}_i^i$, $\tens{t}^{\heart}_i \in \tens{T}^{\heart}_i$, and $\tens{t}^{\spadesuit}_i \in \tens{T}^{\spadesuit}_i$ for arbitrary elements of these sets.
\end{notation}

\Mref{not:union} has the virtue of easing the notation and helping us keep track of different varieties of types, which---as already mentioned---here are IHBs.
Indeed, to interpret \Mref{def:closure} and the corresponding notation just introduced, it is important to stress that the only types appearing in \Mref{def:closure} that do represent IHBs which we---as analysts---assume a player $i$ \emph{actually} to hold are those represented by types in $\tens{T}^{\heart}_i$: all the remaining types, i.e., those in $\tens{T}^{\spadesuit}_i$ and $\tens{T}_{-i}^i$, are IHBs that are a \emph{byproduct} of the implicit context player $i$ considers in her reasoning process.

\begin{example}[continues=ex:CIC, name=Separating Type Structure]
As usual, we focus again on the dynamic game in \Sref{fig:CIC}. Now, if we consider state space $\Omega := S \times \{ t_a \} \times \{t_b\}$ with beliefs described as in \hyperref[tab:belief_function_02]{\color{burgundy}{Table 2}}, for every $i \in I$, we obtain a degenerate and non-common separating type structure. Indeed, by focusing on an arbitrary $i \in I$, we let $\tens{t}^{\heart}_i = t_i$ and $\tens{t}^{\spadesuit}_i = t'_i$, with $\tens{T}^{i}_i := \{ \tens{t}^{\heart}_i \}$ and $\tens{T}^{i}_j := \{ \tens{t}^{\spadesuit}_j \}$, for $j \neq i$. On the contrary, if we focus on the state space $\Omega := S \times \{ t_a \} \times \{t_b\}$ with beliefs described as in the table that follows, we obtain a non-degenerate and common separating type structure.
 
 \begin{table}[H]
\hfill
\parbox{.45\linewidth}{
\centering
\begin{tabular}{@{}rcc@{}} \toprule
	& $\beta^{*}_{a, \la \varnothing \ra} (\overline{t}_a)$ & $\beta^{*}_{a, \la In, Go \ra} (\overline{t}_a)$ \\ 
	\midrule
	$(In\mhyphen Across, t_a)$ & $(0, 1)$ & $(0, 1)$ \\  
	$(Out, t'_a)$ & $(1, 0)$ & $(0,1)$ \\ 
	\bottomrule
\end{tabular}
}
\hfill
\parbox{.45\linewidth}{
\centering
\begin{tabular}{@{}rcc@{}} \toprule
	& $\beta^{*}_{b, \la \varnothing \ra} (\overline{t}_b)$ & $\beta^{*}_{b, \la In \ra} (\overline{t}_b)$ \\ 
	\midrule
	$(Stop, t_b)$ & $(0, 1)$ & $(1, 0)$ \\
	$(Go, t'_b)$ & $(1, 0)$ & $(1, 0)$ \\     
	\bottomrule
\end{tabular}
}
\end{table}

\noindent Indeed, by focusing on an arbitrary $i \in I$, by letting $\tens{t}^{\heart}_i = t_i$ and $\tens{t}^{\spadesuit}_i = t'_i$, we obtain $\tens{T}^{i}_i := \{ \tens{t}^{\heart}_i , \tens{t}^{\spadesuit}_i \}$ and $\tens{T}^{i}_j := \{ \tens{t}^{\heart}_i , \tens{t}^{\spadesuit}_j \}$, for $j \neq i$. Now, incidentally, we can go back to \Mref{rem:existence_uniqueness} and observe that, for the non-belief-closed state space under scrutiny, we can produce two non-trivial separating closures, i.e., separating closures that are \emph{not} the universal state space.
\end{example}

\subsection{Epistemic Events}
\label{subsec:epistemic_events}

To analyze the epistemic events of interest, fix a state space $\widetilde{\Omega}$ and a corresponding separating type structure $\mfrak{T}_i$ for every $i \in I$. As each $\mfrak{T}_i$ is just an instance of a standard type structure with a particular notational convention (from \Mref{not:union} and \Mref{remark:structures_closed_spaces}), we have the same modal operators at our disposal as in \Sref{subsec:interactive_epistemology} and get the the usual epistemic events of $\Rat_i$, $\CSB^{m} (\Rat)$, and $\RCSBR$ as previously introduced within such a type structure. However, rather crucially, these are not the events we are actually interested in, since they merge real and imaginary types. To separate the two, we introduce the following epistemic events, that are the actual focus of this work.

\begin{definition}[Player Dependent Real RmSBR \& Real  RCSBR]
\label{def:player_RCSBR}
Given a state space $\widetilde{\Omega}$, a player $i \in I$, and a corresponding separating type structure $\mfrak{T}_i$, $\CSB^{\heart, m}_i (\Rat) :=  \proj_{i} \CSB^{m} (\Rat) \cap ( S_i \times \tens{T}^{\heart}_i )$ is the event that captures \emph{player $i$'s real RmSBR}, for every $m \in \bN$, with
\begin{equation}
\label{eq:player_real_RCBR}
\RCSBR^{\heart}_i := \CSB^{\heart, \infty}_i (\Rat) := \bigcap_{\ell \geq 0} \CSB^{\heart, \ell}_i (\Rat)
\end{equation}
denoting the event capturing \emph{player $i$'s real RCSBR}.
\end{definition}

It important to  observe that, whereas $\RCSBR^{\heart}_i$ crucially depends on the separating type structure $\mfrak{T}_i$ of player $i$, we keep this dependence notationally implicit.

\begin{remark}[Expressibility]
	For every $i \in I$, $\CSB^{\heart, m}_i (\Rat)$, for every $m \in \bN$, and $\RCSBR^{\heart}_i$ are measurable.\footnote{Note that $\tens{T}^{\heart}_i$ is measurable by assumption and  $\proj_{i} \CSB^{m} (\Rat)$ is closed (see \citet[Footnote 14, p.364]{Battigalli_Siniscalchi_2002}.}
\end{remark}

Finally, we are able to link all the `player dependent' reasoning back together to define the relevant epistemic notion \emph{from the analysts' perspective}.

\begin{definition}[Real RmSBR \& Real  RCSBR]
	\label{def:analyst_RCSBR}
	Given a state space $\widetilde{\Omega}$ and a profile of corresponding separating type structures $(\mfrak{T}_j)_{j \in I}$, $\CSB^{\heart, m} (\Rat) :=  \prod_{j \in I} \CSB^{\heart, m}_j (\Rat)$ is the event that captures \emph{real RmSBR}, for every $m \in \bN$, with
	\begin{equation}
		\label{eq:real_RCBR}
		\RCSBR^{\heart} := \prod_{j \in I} \RCSBR^{\heart}_j
	\end{equation}
	denoting the event that captures \emph{real RCSBR}.
\end{definition}

\section{Characterization}
\label{sec:results}

Clearly, \Mref{rem:characterization} would obtain automatically with a degenerate and common separating type structure. Although, our results cover this case, the interesting and novel parts are the ones about non-degenerate or non-common separating type structures. 

\begin{example}[continues=ex:CIC, name=Non-Degenerate \& Common Separating Type Structure]
Addressing the dynamic game in \Sref{fig:CIC}, as previously done, we focus on the state space $\widetilde{\Omega} := S \times T^\heart$ with $T^{\heart} := \{ t^{\heart}_a \} \times \{ t^{\heart}_b \}$ and beliefs described as in  \hyperref[tab:belief_function_03]{\color{burgundy}{Table 3}} below.
\begin{table}[H]
\hfill
\parbox{.45\linewidth}{
\centering
\begin{tabular}{@{}rcc@{}} \toprule
	& $\beta^{*}_{a, \la \varnothing \ra} (t_a)$ & $\beta^{*}_{a, \la In, Go \ra} (t_a)$ \\ 
	\midrule
	$(Out, t^{\heart}_a)$ & $(0, 1)$ & $(1, 0)$ \\ 
	$(In\mhyphen Across, t^{\heart}_a)$ & $(0, 1)$ & $(1, 0)$ \\   
	$(In\mhyphen Down, t^{\heart}_a)$ & $(0, 1)$ & $(1, 0)$ \\  
	\bottomrule
\end{tabular}
}
\hfill
\parbox{.45\linewidth}{
\centering
\begin{tabular}{@{}rcc@{}} \toprule
	& $\beta^{*}_{b, \la \varnothing \ra} (t_b)$ & $\beta^{*}_{b, \la In \ra} (t_b)$ \\ 
	\midrule
	$(Stop, t^{\spadesuit}_b)$ & $(1, 0, 0)$ & $(0, 1, 0)$ \\
	$(Go, t^{\heart}_b)$ & $(1, 0, 0)$ & $(0, 0, 1)$ \\     
	\bottomrule
\end{tabular}
}
\label{tab:belief_function_03}
\caption{Players' IHBs in state space $\widetilde{\Omega}$.}
\end{table}
\noindent For this state space, consider the non-degenerate and common separating type structure obtained by taking the type spaces $\tens{T}_a$ and $\tens{T}_b$ with beliefs as described in the tables. Within this separating type structure, we have $\RCSBR = \{ (Out, \tens{t}^{\heart}_a) \} \times \{(Go, t^{\heart}_b), (Stop, \tens{t}^{\spadesuit}_b)\}$. Thus, we have that $\proj_S \RCSBR^\heart = \{(Out, Go)\} \notin \FSBRS$.
\end{example}

By the example, it is immediate to observe that FSBRSs fall short on providing a behavioral characterization of non-degenerate and common separating type structures. To find the answer, we have to look at the properties of $\RCSBR^{\heart}$ in common separating type structures.  Now, within a common separating type structure, if $\RCSBR^{\heart}$ is nonempty, then it is nonempty for every player and---by definition---it s a subset of $\RCSBR$, which implies that we necessarily have to get subsets of FSBRSs. However, not all the possible subsets can be obtained, as the \emph{fullness} criterion of Condition (3) in \Mref{def:full_SBRS} is inherited too. To formalize this, we need an additional piece of notation: given a dynamic game $\Gamma$, an FSBRS $F := \prod_{j \in I} F_j \in \FSBRS$, a player $i \in I$, and a strategy $s^*_i \in F_i$, we let $\mu_i^{s^*_i, F} \in \Delta^{\mcal{H}_i} (S_{-i})$ denote a CPS that justifies $s^{*}_i$ within $F$, i.e., a CPS satisfying Condition (1) and (2) in \Mref{def:full_SBRS}.

\begin{definition}[Misaligned FSBRS]
	\label{def:MFSBRS}
	Given a dynamic game $\Gamma$, a product set $M := \prod_{j \in I} M_j \subseteq S$ is a \emph{misaligned FSBRS} (henceforth, MFSBRS) if there exists an $F = \prod_{j \in I} F_j \in \FSBRS$ such that for every $i \in I$ and every $s_i^* \in M_i$:
	\begin{enumerate}[label=\arabic*)]
		\item $s_i^* \in F_i$, and
		
		\item $s_i \in M_i$, for every $s_i \in \rho_i \big( \mu_i^{s_i^*, F} \big)$,
	\end{enumerate}
where $\mu_i^{s_i^*, F}$ is as defined above and it is well-defined by Condition (1).
\end{definition}

As it has been for \Mref{def:full_SBRS} and \Mref{def:player-specific_FSBRS}, we now provide the player-specific counterpart of \Mref{def:MFSBRS}.

\begin{definition}[Player-Specific MFSBRSs]
\label{def:player-specific_MFSBRS}
Given a dynamic game $\Gamma$ and a player $i \in I$, a set $M_i \subseteq S_i$ is a \emph{player $i$'s specific MFSBRS} if there exists an $M_{-i} := \prod_{j \in I \setminus \{i\}} M_j$ such that $M_i \times M_{-i}$ is an MFSBRS.
\end{definition} 

In the following, given that we let $\MFSBRS$ denote the family of MFSBRSs of a dynamic game $\Gamma$, it is immediate to observe that the family of player $i$'s specific MFSBRSs, denoted $\MFSBRS_i$, is captured by  
\begin{equation*}
\MFSBRS_i := \Set { M_i \in \wp (S_i) | \exists M \in \MFSBRS : \proj_i M = M_i },
\end{equation*}
for every $i \in I$. As it is for the case of FSBRSs (as pointed out in \Sref{subsec:solution_concepts}), it is not necessarily the case that $\prod_{j \in I} M_j \in \MFSBRS$. In what follows, we address this issue along with how MFSBRSs look like in general in our running example.

\begin{example}[continues=ex:CIC, name=MFSBRS]
For our running example, it is easy to see that 
\begin{equation*}
\MFSBRS = \{ 
  			\{ In\mhyphen Across\} \times \{ Go\} , %
  			\{ Out \} \times \{Stop \} ,%
  			\{ Out \} \times \{Go \}, %
  			\{ Out \} \times \{Stop,Go \}, %
  			\{ \emptyset \} %
\},
\end{equation*}
with the player-specific MFBSBRs given by
\begin{align*}
\MFSBRS_a & = \{ %
\{ In\mhyphen Across \}  , %
\{ Out \},
\{ \emptyset \}
\} ,\\
\MFSBRS_b & = \{ %
\{ Stop \} , %
\{ Go \}, %
\{ Stop, Go \} , %
\{ \emptyset \}
\} .
\end{align*}
\noindent Relative to $\FSBRS$, one new set appears, namely, $\{ Out \} \times \{Go \}$, which clearly shows some misalignment, because, if Ann would indeed believe that Bob plays $Go$, she should not play $Out$. Also, it is immediate to observe that $\{ In \mhyphen Across \} \times \{ Stop \} \notin \MFSBRS$, whereas $\{ In \mhyphen Across \} \in \MFSBRS_a$ and $\{ Stop \} \in \MFSBRS_b$.
\end{example}

Now, we are ready to state the main result of our paper, which links the notion of Real RCSBR in separating type structures to those of MFSBRSs and FSBRSs.

\begin{proposition}
\label{prop:characterization_static_1}
Given a dynamic game $\Gamma$, a state space $\widetilde{\Omega}$, and a profile of corresponding separating type structures $(\mfrak{T}_j)_{j \in I}$, the following hold.
\begin{enumerate}[label=\arabic*)]
	\item Given the profile $(\mfrak{T}_j)_{j \in I}$, $\proj_{S} \RCSBR^{\heart} \in \prod_{j \in I} \MFSBRS_j$.
	
	\item If $(\mfrak{T}_j)_{j \in I}$ is common, then $\proj_{S} \RCSBR^{\heart} \in \MFSBRS$.
	
	\item If $\mfrak{T}_j$ is degenerate for every $j \in I$, then $\proj_{S} \RCSBR^{\heart} \in \prod_{j \in I} \FSBRS_j$.
	
	\item If $(\mfrak{T}_j)_{j \in I}$ is common and degenerate, then $\proj_{S} \RCSBR^{\heart} \in \FSBRS$.
\end{enumerate}
\end{proposition}

The results of \Mref{prop:characterization_static_1} are concisely summarized in \hyperref[tab:characterization]{\color{burgundy}{Table 4}}: for example, the entry corresponding to ``non-degenerate'' and ``non-common'' corresponds to Part (1) of \Mref{prop:characterization_static_1}.

\begin{table}
\renewcommand{\arraystretch}{1.2}
\centering
\begin{tabular}{@{}lcc@{}} \toprule
					& common 					& non-common \\ \midrule
degenerate     		&   $\FSBRS$     				& $\prod_{j \in I} \FSBRS_j$     \\
non-degenerate 	&   $\MFSBRS$     & $\prod_{j \in I} \MFSBRS_j$       \\ \bottomrule
\end{tabular}
\label{tab:characterization}
\caption{Behavioral implications of real RCSBR with varying assumptions about separating type structures.}
\end{table}

The next two examples illustrate that \Mref{prop:characterization_static_1} is tight, in the sense that, starting from the assumptions in Part (2) (resp., Part ($3$)), we cannot get the stronger conclusion of Part ($3$) (resp., Part ($4$)). In doing so, incidentally, they also show how the outcomes $\{ Out \} \times \{ Go \}$ and $\{ In\mhyphen Across \} \times \{ Stop \}$ are indeed compatible with RCSBR when we allow for misalignment of beliefs.

\begin{example}[continues=ex:CIC, name=Non-Degenerate \& Common Separating Type Structure]
We go back to the---already covered at the beginning of this section---case of a non-degenerate and common separating type structure by addressing the dynamic game in \Sref{fig:CIC} with state space  $\widetilde{\Omega} := S \times T^\heart$ and beliefs described as in \hyperref[tab:belief_function_03]{\color{burgundy}{Table 3}}. We already observed that in this case we have $\proj_S \RCSBR^\heart = \{(Out, Go)\} \notin \FSBRS$. However, we do have that $\proj_S \RCSBR^\heart = \{(Out, Go)\} \in \MFSBRS$, as it should be according to  Part (2) of \Mref{prop:characterization_static_1}.
\end{example}

\begin{example}[continues=ex:CIC, name=Degenerate \& Non-Common Separating Type Structure]
We study again the dynamic game in \Sref{fig:CIC} by focusing on the state space $\widetilde{\Omega}' := S \times T'^\heart$ with $T'^{\heart} := \{ t^{\heart}_a \} \times \{ t^{\heart}_b \}$ and types' beliefs as described in the table below, which is exactly like \hyperref[tab:belief_function_02]{\color{burgundy}{Table 2}} with $t^{\heart}_i = t_i$ and $t^{\spadesuit}_i = t'_i$ for every $i \in I$.
 \begin{table}[H]
\hfill
\parbox{.45\linewidth}{
\centering
\begin{tabular}{@{}rcc@{}} \toprule
	& $\beta^{*}_{a, \la \varnothing \ra} (t_a)$ & $\beta^{*}_{a, \la In, Go \ra} (t_a)$ \\ 
	\midrule
	$(In\mhyphen Across, t^{\heart}_a)$ & $(0, 0, 1)$ & $(0, 0, 1)$ \\  
	$(Out, t^{\spadesuit}_a)$ & $(1, 0, 0)$ & $(0,1, 0)$ \\ 
	$(In\mhyphen Down, t^{\spadesuit}_a)$ & $(1, 0, 0)$ & $(0,1, 0)$ \\  
	\bottomrule
\end{tabular}
}
\hfill
\parbox{.45\linewidth}{
\centering
\begin{tabular}{@{}rcc@{}} \toprule
	& $\beta^{*}_{b, \la \varnothing \ra} (t_b)$ & $\beta^{*}_{b, \la In \ra} (t_b)$ \\ 
	\midrule
	$(Stop, t^{\heart}_b)$ & $(0, 1, 0)$ & $(0, 0, 1)$ \\
	$(Go, t^{\heart}_b)$ & $(0, 1, 0)$ & $(0, 0, 1)$ \\
	$(Go, t^{\spadesuit}_b)$ & $(1, 0, 0)$ & $(1, 0, 0)$ \\     
	\bottomrule
\end{tabular}
}
\end{table}
\noindent For this state space, define Ann's separating type structure by setting $\tens{T}^a_a := \{ \tens{t}^{\heart}_a\}$ and $\tens{T}^a_b := \{ \tens{t}^{\spadesuit}_b\}$, whereas for Bob let $\tens{T}^b_a := \{ \tens{t}^{\spadesuit}_a \}$ and $\tens{T}^b_b := \{ \tens{t}^{\heart}_b\}$, with beliefs as given above. Note that both separating type structures are degenerate and non-common and according to Part (3) of \Mref{prop:characterization_static_1} we should get $\proj_S \RCSBR^{\heart}_i \in \FSBRS_i$. Indeed, it is immediate to verify that for Ann's separating type structure, we have $\proj_S \RCSBR^{\heart}_a = \{ In\mhyphen Across \}$, whereas for Bob's we get $\proj_S \RCSBR^{\heart}_b = \{ Stop \}$. Incidentally, this state space produces the outcome of the game mentioned in the introduction.
\end{example}

The previous result is tight also in another sense: we have the usual `epistemic characterization converse', which is in the spirit of Part (2) of \Mref{rem:characterization} and delivers---along with \Mref{prop:characterization_static_1}---the characterization we are after.

\begin{proposition}
\label{prop:characterization_static_2}
Given a dynamic game $\Gamma$, the following hold.
\begin{enumerate}[label=\arabic*)]
	\item Let $M_i \in \MFSBRS_i$ be nonempty, for every $i \in I$. Then there exists a state space and a corresponding profile of finite separating type structures $(\mfrak{T}_j)_{j \in I}$ such that $\proj_{S} \RCSBR^{\heart} = \prod_{j \in I} M_j$.
	
	\item Let $M \in \MFSBRS$ be nonempty. Then there exists a state space and a corresponding finite common separating type structure $\mfrak{T}$ such that $\proj_{S} \RCSBR^{\heart} = M$.
	
	\item Let $F_i \in \FSBRS_i$ be nonempty, for every $i \in I$. Then there exists a state space and a corresponding profile of finite and degenerate separating type structures $(\mfrak{T}_j)_{j \in I}$ such that $\proj_{S} \RCSBR^{\heart} = \prod_{j \in I} F_j$.
	
	\item Let $F \in \FSBRS$ be nonempty. Then there exists a state space and a corresponding finite, degenerate, and common separating type structure $\mfrak{T}$ such that $\proj_{S} \RCSBR^{\heart} = F$.
\end{enumerate}
\end{proposition}

\section{Non-Monotonic vs. Monotonic Reasoning}
\label{sec:non-monotonicity}

It is immediate to establish that \Mref{prop:characterization_static_1} and \Mref{prop:characterization_static_2} cover static games when the focus is on the implications of \emph{Rationality and Common Belief in Rationality}. However, there is a significant difference between static games and dynamic games, whose origin has to be found in the different nature of reasoning that we can have in these classes of games when the focus is on the usage of the belief operator. In the remainder of this section we clarify this point.

Thus, first of all, as we point out in \Sref{subsec:motivation_results}, our focus on dynamic games and on type structures suited for their analysis allows us to easily compare our results to the corresponding results in the context of static games, since the latter are naturally a degenerate case of the representation we employ with  $H^{\varnothing}_i = \mcal{H}_i = \{ \la \varnothing \ra \}$, for every $i \in I$. Thus, in particular, regarding interactive epistemology, we have that $\beta_i := \beta_{i, \la \varnothing \ra}$ and $\Bel_i := \Bel_{i, \la \varnothing \ra}$, for every $i \in I$, with $\Bel$ canonically defined, which allows us to introduce the \emph{correct belief} operator $\CB (E) := E \cap \Bel (E)$, for every $E \in \mcal{A} (\Omega)$. This in turn leads to $\CB^{m} (\Rat)  :=  \Rat \cap \bigcap^{m - 1}_{k = 0} \Bel (\CB^{k} (\Rat))$ and, finally, to $\RCBR  := \CB^{\infty} (\Rat)  :=  \bigcap_{\ell \geq 0} \CB^\ell (\Rat)$, which captures \emph{Rationality and Common Belief of Rationality} (henceforth, RCBR), where---naturally, given what is written above concerning best-replies---the notion of rationality for static games is a `degenerate' version of the one for dynamic games. Now, \citet[Proposition 2.1, p.1396]{Brandenburger_Dekel_1987} show\footnote{The original formulation in \citet[Proposition 2.1, p.1396]{Brandenburger_Dekel_1987} is based on a different framework from the one employed here, which is the one of \citet[Theorem 12.1, p.635]{Dekel_Siniscalchi_2015}. To see the relation between the two, see \citet[Theorem 12.3, p.639]{Dekel_Siniscalchi_2015}.} that, given a static game $\Gamma$:
\begin{enumerate}[label=\arabic*)]
\item for every standard type structure $\mscr{T}$, $\proj_S \RCBR$ is a Full Best-Reply Set (henceforth, FBRS);

\item for every FBRS $F$, there exists a finite standard type structure $\mscr{T}$  such that $\proj_S \RCBR = F$,
\end{enumerate}
where a FBRS is a FSBRS as in \Mref{def:full_SBRS}, with the---obvious---caveat that Condition (2) boils down to belief, given the static nature of the setting.
Also, \citet[Theorems 5.1 \& 5.3, p.378 \& p.379]{Tan_Werlang_1988} show that $\proj_S \RCBR = \PR^\infty$ in $\mscr{T}^*$, where $\PR^{\infty}$ is defined as the \emph{set of correlated rationalizable strategy profiles} of \cite{Bernheim_1984} and \cite{Pearce_1984} (see \citet[Definition 54.1, Chapter 4.1]{Osborne_Rubinstein_1994}).\footnote{And, of course, it coincides with Strong Rationalizability, whenever the latter is applied on a static game (considered as a degenerate dynamic game).}

The crucial aspect of the results described in the paragraph above is that in the static case we always have that $F \subseteq \PR^\infty$, for every FBRS $F$, contrary to what happens with Strong Rationalizability and FSBRSs, as illustrated  in \Mref{ex:CIC}(FSBRSs) in \Sref{subsec:solution_concepts}\footnote{\cite{Battigalli_Friedenberg_2012} use the dynamic game known as ``Battle of the Sexes with an Outside of Option'' to show non-monotonic reasoning. The reader is referred to their exposition and, in particular, to their Examples 3 and 4 at pages 63--64, for more details about non-monotonicity.} and recorded in \Mref{rem:SR_FSBRS}. This is the result of the fact that Rationalizability is built on the notion of belief (and, in an epistemic framework, on the operator $\Bel_i$), which is---as pointed out in \Sref{subsec:beliefs}---monotonic, whereas Strong Rationalizability is based on the notion of strong belief (and, in an epistemic framework, on the operator $\SB_i$), which does \emph{not} satisfy monotonicity---again as pointed out in \Sref{subsec:beliefs}. Thus, adapting our characterizations (i.e., \Mref{prop:characterization_static_1} and \Mref{prop:characterization_static_2}) to the static case, we would not get any further predictions beyond $\PR^\infty$, even when allowing players to have misaligned beliefs. However, the tightness of our result carries over to static games too, as illustrated next.

\begin{example}[label=ex:counterexample, name=Static game]
We now focus on the static game in \Sref{fig:counterexample}. It can be easily verified that $\{U\} \times \{R\}$ is not a FBRS.
	
	\begin{figure}[H]
		\centering
		\begin{tikzpicture}
			[info/.style={circle, draw, inner sep=1.5, fill=black},
			scale=1.1] 
			\draw (0, 0) node {};
			\draw (-3.5, 0) node {
				\begin{game}{3}{3}[Ann][Bob]
					& $L$ 		& $C$ 		& $R$	\\
					$U$ 	& $1 ,2$ 	& $2, 1$ 	& $0, 0$ \\
					$M$ 	& $0, 0$ 	& $1, 2$	& $2, 1$ \\
					$D$ 	& $2, 1$ 	& $0, 0$ 	& $1, 2$\\
			\end{game}};
		\end{tikzpicture}
		\caption{A static game.}
		\label{fig:counterexample}
	\end{figure}
	
\noindent We let $T_i := \{ t^{\heart}_i , t^{\spadesuit}_i, t'^{\spadesuit}_i \}$, for every $i \in I$, with beliefs described by the following tables.
	
\begin{table}[H]
\parbox{.45\linewidth}{
\centering
		\begin{tabular}{@{}rc@{}} \toprule
			& $\beta^{*}_{a} (t_a)$ \\ \midrule
			$(U, t^{\heart}_a)$ &  $(0, 1, 0)$   \\  
			$(M, t^{\spadesuit}_a)$ & $(0, 0, 1)$   \\ 
			$(D, t'^{\spadesuit}_a)$ &  $(1, 0, 0)$  \\  \bottomrule
		\end{tabular}
}
\hfill
\parbox{.45\linewidth}{
\centering
		\begin{tabular}{@{}rc@{}} \toprule
				& $\beta^{*}_{b} (t_b)$  \\ \midrule
				$(L, t^{\spadesuit}_b)$  & $(1, 0, 0)$  \\  
				$(C, t'^{\spadesuit}_b)$  & $(0, 1, 0)$  \\ 
				$(R, t^{\heart}_b)$  & $(0, 0, 1)$  \\  
				\bottomrule
			\end{tabular}
}
\end{table}
		
\noindent Now, we focus on the state space $\widetilde{\Omega}'' := S \times T''^{\heart}$, with $T''^{\heart}_i := \{ t^{\heart}_i \}$ for every $i \in I$, where---following our convention---we have that $t^{\heart}_i \in T''^{\heart}_i$ is the (only) actual IHB that player $i$ holds, for every $i \in I$. Taking the player-specific closure $\mfrak{T}_i$, for every $i \in I$, we obtain the non-degenerate and common separating type structure with $\tens{T}_i := \{ \tens{t}^{\heart}_i, \tens{t}^{\spadesuit}_i, \tens{t}'^{\spadesuit}_i \}$, for every $i \in I$. In this common separating type structure, we have that all types belong to $\RCBR$. As a result, we have that $\proj_{S_a} \RCBR^{\heart}_a = \{ U \}$ and  $\proj_{S_b} \RCBR^{\heart}_b = \{ R \}$. However, note that---in contrast to our leading example---here $\proj_{S_a} \RCBR^{\heart}_i \subseteq \PR^\infty$, since under RCBR there is no issue with monotonicity.
\end{example}

In the example above, we only consider a \emph{common} separating type structure, which, in a sense, is without loss of generality when there is no issue of non-monotonicity. We now clarify this crucial point. Under strong belief, we need to introduce player-specific closures to disentangle the players' reasoning from the assumptions made by the analyst: without these player-specific closures, players would be forced to reason about IHBs that only the analyst actually considers. Here, with static games and RCBR, we could instead dismiss player-specific closures, because we can always take the union across all the player-specific separating type structures to obtain one common separating type structure. Since (common) belief is monotonic, this union would not alter the way in which players reason and, in particular, it would not force them to reason in a specific manner either. Without monotonicity, such a union of type structures would on the contrary alter the reasoning. 

Finally, it is important to clarify why we state above that the difference between static and dynamic games arises when the focus is on the usage of the belief operator. The reason lies in the nature of the \emph{assumption} operator as typically defined in the literature on the epistemic analysis of Iterated Admissibility stemming from \cite{Brandenburger_et_al_2008}, hence, in the study of static games. Crucially, this modal operator presents a form of non-monotonic reasoning much in the same spirit of the strong belief operator, as emphasized in \citet[Section 8.3]{Brandenburger_Friedenberg_2010}. Thus, the difference is \emph{not} in the class of games \emph{per se}, but rather on the tools employed to perform the epistemic analysis.

\section{Discussion}
\label{sec:discussion}

\subsection{On Player-Specific Type Structures}
\label{subsec:player_specfic_structures}

\citet[Section 8.1]{Brandenburger_Friedenberg_2010} propose to study player-specific type structures, where type structures are introduced for every player with the understanding that  they can be potentially different. The idea behind these objects  is to capture lack of transparency between the players involved in a strategic interactions of the beliefs that are possible in a given type structure. As a matter of fact, our notions of separating closure (as in \Mref{def:closure}) and of separating type structure (as in \Mref{def:separating_type_structure}) can be interpreted as generalizations of this idea in a parsimonious way. Indeed, if we have a profile of separating type structures $(\mfrak{T}_j)_{j \in I}$, where $\mfrak{T}_i$ is degenerate for every $i \in I$, then we effectively have what they call player-specific type structures. This allows us to answer\footnote{The formal answer is given by Part (3) in our \Mref{prop:characterization_static_1} and \Mref{prop:characterization_static_2} when interpreted for RCBR in static games. See also \Sref{sec:non-monotonicity}.} in the affirmative to a conjecture made in \citet[Section 8.1, p.802]{Brandenburger_Friedenberg_2010}, where it is stated that, contrary to non-monotonic reasoning, monotonic reasoning in its behavioral prediction is somewhat unaffected by lack of transparency. Indeed, interpreting our characterizations (\Mref{prop:characterization_static_1} and \Mref{prop:characterization_static_2}) for the static case as done in  \Sref{sec:non-monotonicity}, our analysis  shows that, even allowing for more misalignment, the behavioral predictions for static games under RCBR never fall outside those provided by $\PR^\infty$.

\subsection{On the Notion of State Space}
\label{subsec:state_space}

Our definition of state space does \emph{not} allow to restrict actual play. In principle, one could allow for such restrictions too, which are somewhat implicit in  epistemic models \emph{\`{a} la} \cite{Aumann_1987}. However, we refrain from doing so, because our interest lies in an analyst that considers certain players' IHBs and that wants to characterize the behavioral implications for these IHBs satisfying certain conditions. Thus, restricting play upfront would be against the very nature of our analysis.\footnote{For a related point, see \citet[Footnote 5, p.35]{Stalnaker_1998}.}

\subsection{$\Delta$-Restrictions}
\label{subsec:Delta}

As it is well known from \citet[Section 6]{Battigalli_Friedenberg_2012}, the relation between contextual assumptions as captured via  standard type structures and FSBRSs can be also captured via so-called $\Delta$-restrictions. Those are explicit restrictions on players' first order beliefs: in particular, the $\Delta$ in the name stands for a profile  $\Delta := (\Delta_j)_{j \in I}$, where  $\Delta_i \subseteq \Delta^{\mcal{H}_i} (S_{-i})$, for every $i \in I$. Crucially, $\Delta$-restrictions are assumed to be transparent between the players.  Hence, they cannot play any role in the present work, because by allowing for non-belief-closed state spaces such a transparency between the players might be lacking. However, with the equivalence of \citet[Section 6]{Battigalli_Friedenberg_2012} in mind, starting with a profile of separating type structures $(\mfrak{T}_j)_{j \in I}$ one could associate them with player-specific $\Delta$-restrictions $(\Delta^j)_{j \in I}$. The interpretation then would be that player $i$ is certain that the $\Delta$-restriction $\Delta^i$ is transparent between all the players, but she might be wrong whenever there exists a  $j \in I \setminus \{i\}$ with $\Delta^j \neq \Delta^i$.

\appendix

\section*{Appendix}

\section{Proofs}
\label{app:proofs}


\begin{proof}[Proof of \Mref{prop:characterization_static_1}]
	Fix a dynamic game $\Gamma$, a state space $\widetilde{\Omega}$, and a profile of corresponding separating type structures $(\mfrak{T}_j)_{j \in I}$.
	
\begin{enumerate}[leftmargin=*, label=\arabic*)]

		\item If there exists an $i \in I$ such that $\RCSBR^{\heart}_i$ is empty, then  the result follows immediately. %
		Thus, assume that $\RCSBR^{\heart}_i$  is nonempty for every $i \in I$. %
		Hence, given a player $i \in I$ with separating type structure $\mfrak{T}_i$, from \Mref{def:player_RCSBR} we have that  $\proj_j \RCSBR$  relative to $\mfrak{T}_i$ is nonempty, for every $j \in I$. %
		 Now, from Part (1) of \Mref{rem:characterization}, this in turn implies that for every $i \in I$ with separating type structure $\mfrak{T}_i$ there exists an $F^i := \prod_{j \in I} F^i_j \in \FSBRS$, with $F^i_j \in \FSBRS_j$ for every $j \in I$, such that $\proj_S \RCSBR = F^i$ relative to $\mfrak{T}_i$. This, in turn, induces the existence of an $(F^j)_{j \in I} \in \FSBRS^I$. %
		 Now, for every $i \in I$ with separating type structure $\mfrak{T}_i$, let $M_i := \proj_{S_i} \RCSBR^{\heart}_i$, with $\RCSBR^{\heart}_i$ obtained relative to the separating type structure $\mfrak{T}_i$, and note that, by \Mref{def:analyst_RCSBR}, we have $M := \prod_{j \in I} M_j = \proj_S \RCSBR^{\heart}$. %
		 Thus, it remains to show that $M_i \in \mathcal{M}_i$, for every $i \in I$. Regarding Condition (1) of \Mref{def:MFSBRS}, since $\RCSBR^{\heart}_i \subseteq \RCSBR_i$, we have that $M_i \subseteq F^i_i$. Finally, the fullness Condition (2) of \Mref{def:MFSBRS} is inherited from $F_i^i \in \FSBRS_i$ as part of the FSBRS $F^i$.
		
		\item Assume that $(\mfrak{T}_j)_{j \in I}$ is common, which allows to set $\mfrak{T} := (\mfrak{T}_j)_{j \in I}$.  Following the arguments used to prove Part (1), we obtain a profile $(F^j)_{j \in I} \in \FSBRS^I$, which we simply refer to as $F := \prod_{j \in I} F_j \in \FSBRS$ in light of the fact that $F^i = F^j$ for every $i, j \in I$ is an immediate consequence of $\mfrak{T}$ being a common separating type structure. Now, given that we let $M := \proj_S \RCSBR^{\heart}$, it remains to show that $M \in \MFSBRS$. Since $\RCSBR^{\heart}_i \subseteq \RCSBR_i$, we have that $M_i \subseteq F_i$, establishing Condition (1) of \Mref{def:MFSBRS}. Finally, the fullness Condition (2) of \Mref{def:MFSBRS} is inherited from $F_i \in \FSBRS_i$ as part of the FSBRS $F$.
		
		\item Suppose $\mfrak{T}_i$ is degenerate for every $i \in I$. Thus, from Part (1) of \Mref{rem:characterization}, we know that for every $i \in I$, $\proj_S \RCSBR \in \FSBRS$ within the separating  type structure $\mfrak{T}_i$. Furthermore, since $\mfrak{T}_i$ is degenerate, we have that $\RCSBR = \RCSBR^\heartsuit$.  Hence, relative to the same separating type structure we have $\proj_{S_i} \RCSBR^\heartsuit \in \FSBRS_i$ and the result follows from \Mref{def:analyst_RCSBR}.
		
		\item If $(\mfrak{T}_j)_{j \in I}$ is common and degenerate, the result follows directly from Part (1) of \Mref{rem:characterization}.  \qedhere
	\end{enumerate}
\end{proof}

\begin{proof}[Proof of \Mref{prop:characterization_static_2}]
	Fix a dynamic game $\Gamma$.
	
	\begin{enumerate}[leftmargin=*, label=\arabic*)]
	
\item Suppose $M_i \in \MFSBRS_i$ for every $i \in I$. By definition, for every $i \in I$ there exists a profile $(F^i_j)_{j \in I} \subseteq S$ such that: ($1$) $M_i \subseteq F^i_i$; ($2$) $F^i_j \in \FSBRS_j$, for every $j \in I \setminus \{i\}$; ($3$) $F^i:=F^i_i \times F^i_{-i} \in \FSBRS$; ($4$) $F^i_i$ satisfies Condition (3) of \Mref{def:full_SBRS} relative to  $F^i$. Let
		\begin{equation*}
			\mscr{T}_i := \la (T^i_i, \beta^i_i), (T^i_j , \beta^i_j )_{j \in I \setminus \{i\}} \ra 
		\end{equation*}
		denote the corresponding finite standard type structure that satisfies Part (2) in \Mref{rem:characterization}, i.e., such that $\proj_S \RCSBR = F^i$ relative to $\mscr{T}_i$. 
		Now, we are ready to construct a separating type structure $\mfrak{T}_i$ for every $i \in I$ as follows, with the separating closure of the state space as an immediate consequence of this construction.
		
		\begin{itemize}[leftmargin=*]
			\item \emph{Type spaces:} For player $i \in I$, recalling that given our convention we always have $T_i^i \subseteq T_i^*$, we let
			\begin{equation*}
				\tens{T}^{\heart, i}_i := \Set {  t_i^* \in T_i^i | \exists s_i \in M_i :  %
					s_i \in \rho_i (\marg_{S_{-i}} \beta^i_i(t_i^*))  }
			\end{equation*}
			and $\tens{T}^{\spadesuit, i}_i := T_i^i \setminus \tens{T}^{\heart,i}_i$.
		 For $j \in I \setminus \{i\}$, let $\tens{T}^{\heart,i}_j := T^i_j$ and  $\tens{T}^{\spadesuit,i}_j := \emptyset$, where it has to be observed that this construction---naturally---involves non-redundant types only.

			\item \emph{Belief functions:} For every $i, j \in I$ and $\tens{t}_j \in \tens{T}^i_j$, we let  $\bm{\beta}^i_j (\tens{t}_j) := \beta_j^*(\tens{t}_j)$.
		\end{itemize}
		
		\noindent Now, for this part it remains to prove that we have $M_i = \proj_{S_i} \RCSBR_i^{\heart}$ relative to the separating type structure $\mfrak{T}_i$ of player $i \in I$ just constructed. First, consider $s_i \in M_i$. By construction, there exists $t_i^* \in \tens{T}^{\heart,i}_i$ such that $s_i \in \rho_i (\marg_{S_{-i}} \beta^i_i (t_i^*))$. Furthermore, we have that $t_i^* \in T_i^i$, implying that $(s_i, t_i^*) \in \proj_{i} \CSB^{m} (\Rat)$ for $m\geq 1$ and $m=\infty$, which in turn implies that $(s_i, t_i^*) \in \proj_{i} \CSB^{\heart, m} (\Rat)$ for $m \geq 1$. Therefore, we have that $(s_i, t_i^*) \in \RCSBR^{\heart}_i$.
		
		Conversely, consider $(s_i, t_i^*) \in \RCSBR^{\heart}_i\subseteq \proj_i \RCSBR$. Thus, $s_i \in F_i^i$. Since, 
\begin{equation*}
t_i^* \in \proj_{\tens{T}_i} \RCSBR^{\heart}_i \subseteq \tens{T}^{\heart,i}_i, 
\end{equation*}
there exists a $s_i^* \in M_i \subseteq F_i^i$ such that $s_i^* \in \rho_i (\marg_{S_{-i}} \beta^i_i(t_i^*))$. If $s_i^* = s_i$ we are done. If not, using the notation as introduced immediately before \Mref{def:MFSBRS}, note that there exists a $\mu_i^{s_i^*, F^i}$ such that $\marg_{S_{-i}} \beta_i(t_i^*) = \mu_i^{s_i^*, F^i}$. Hence, since $(s_i, t_i^*) \in \Rat^{\heart}_i$, we also have that $s_i \in \rho_i (\mu_i^{s_i^*, F^i})$, which, thanks to Condition (2) of \Mref{def:MFSBRS}, ensures that $s_i \in M_i$.

\item Assume that $M \in \MFSBRS$. Thus, for every $i$ and $i^\prime \in I$, we can take $(F^i_j)_{j \in I}=(F^{i^\prime}_j)_{j \in I}$ in the previous construction and  \emph{one} finite standard type structure $\mscr{T} := \la (T_j , \beta_j )_{j \in I} \ra$ as the one that satisfies Part (2) in \Mref{rem:characterization}, i.e., such that $\proj_S \RCSBR = F:=\prod_{j \in I} F^i_j$ for one (and all) $i \in I$ relative to $\mscr{T}$. Now, we are ready to construct a common separating type structures $\mfrak{T}$ as follows, with the separating closure of the state space as an immediate consequence of this construction.
\begin{itemize}[leftmargin=*]
\item \emph{Type spaces:} For every $i \in I$, recalling that given our convention we always have $T^i \subseteq T_i^*$, we let
\begin{equation*}
	\tens{T}^{\heart}_i := \Set {  t_i^* \in T_i | \exists s_i \in M_i :  %
		s_i \in \rho_i (\marg_{S_{-i}} \beta_i(t_i^*))  }
\end{equation*}
and $\tens{T}^{\spadesuit}_i := T_i \setminus \tens{T}^{\heart}_i$.

\item \emph{Belief functions:} For every $j \in I$ and $\tens{t}_j \in \tens{T}_j$, we let  $\bm{\beta}_j (\tens{t}_j) := \beta_j^*(\tens{t}_j)$.
\end{itemize}
The rest of the proof of this part obtains exactly from the same steps employed to establish Part (1).
		
\item  If $F_i \in \FSBRS_i$ for every $i \in I$, then it has to be observed that, from the construction employed to establish Part (1), we get 
$\tens{T}^{\spadesuit,i}_i = \emptyset$ for every $i \in I$. Hence, this makes all the constructed separating type structures degenerate.

\item If $F \in \FSBRS$ the result follows directly from Part (2) in \Mref{rem:characterization}, because the corresponding standard type structure obtained there can be seen as a finite, degenerate, and common separating type structure by appropriate labeling. \qedhere
\end{enumerate}
\end{proof}

%
\hypersetup{colorlinks=true,linkcolor=green!50!black}
\phantomsection
\addcontentsline{toc}{section}{References}


\bibliography{./Background/Biblio_Players_Specific_Type_Structures.bib}{}


\bibliographystyle{./Background/ampersand_standard_capital_natbib}

\end{document}